\documentclass[twocolumn]{aastex631} %, linenumbers

\usepackage{CJK}
\usepackage{times}
\usepackage{amsmath}
\usepackage{graphicx}
\usepackage{hyperref}
\usepackage{gensymb}
\usepackage{upgreek}
\usepackage{natbib}
\usepackage{subfigure}
\usepackage{multirow}
\usepackage{comment}
\usepackage{mathtools}
\usepackage{mathrsfs}
\usepackage{fontenc}
\usepackage{color}
\usepackage{url}

\begin{document}
\begin{CJK*}{UTF8}{gbsn}

\title{High-quality Extragalactic Legacy-field Monitoring (HELM) with DECam}

%\red{(acronym is open to suggestions)}

\author[0000-0001-5105-2837]{Ming-Yang Zhuang (庄明阳)}
\email{mingyang@illinois.edu}
\affiliation{Department of Astronomy, University of Illinois Urbana-Champaign, Urbana, IL 61801, USA}

\author[0000-0002-6893-3742]{Qian Yang}
\affiliation{Center for Astrophysics $\vert$ Harvard \& Smithsonian, 60 Garden Street, Cambridge, MA 02138, USA}
\affiliation{Department of Astronomy, University of Illinois Urbana-Champaign, Urbana, IL 61801, USA}

\author[0000-0003-1659-7035]{Yue Shen}
\affiliation{Department of Astronomy, University of Illinois Urbana-Champaign, Urbana, IL 61801, USA}
\affiliation{National Center for Supercomputing Applications, University of Illinois Urbana-Champaign, Urbana, IL 61801, USA}

\author[0000-0002-6904-359X]{Monika Adam\'ow}
\affiliation{Center for AstroPhysical Surveys, National Center for Supercomputing Applications, University of Illinois Urbana-Champaign, Urbana, IL 61801, USA}

\author[0000-0002-3632-7668]{Douglas N. Friedel}
\affiliation{National Center for Supercomputing Applications, University of Illinois Urbana-Champaign, Urbana, IL 61801, USA}

\author[0000-0002-4588-6517]{R.~A. Gruendl}
\affiliation{National Center for Supercomputing Applications, University of Illinois Urbana-Champaign, Urbana, IL 61801, USA}

\author[0000-0003-0049-5210]{Xin Liu}
\affiliation{Department of Astronomy, University of Illinois Urbana-Champaign, Urbana, IL 61801, USA}
\affiliation{National Center for Supercomputing Applications, University of Illinois Urbana-Champaign, Urbana, IL 61801, USA}
\affiliation{Center for Artificial Intelligence Innovation, University of Illinois at Urbana-Champaign, 1205 West Clark Street, Urbana, IL 61801, USA}

\author[0000-0002-4279-4182]{Paul Martini}
\affiliation{Department of Astronomy, The Ohio State University, Columbus, Ohio 43210, USA}
\affiliation{Center of Cosmology and Astro-Particle Physics, The Ohio State University, Columbus, Ohio 43210, USA}

% alphabetical order below

\author[0000-0003-1587-3931]{Timothy M. C. Abbott}
\affiliation{CTIO, NSF’s NOIRLab, Casilla 603, La Serena, Chile}

\author[0000-0002-6404-9562]{Scott F. Anderson}
\affiliation{Astronomy Department, University of Washington, Box 351580, Seattle, WA 98195, USA}

\author[0000-0002-9508-3667]{Roberto J. Assef}
\affiliation{N\'ucleo de Astronom\'ia de la Facultad de Ingenier\'ia y Ciencias, Universidad Diego Portales, Av. Ej\'ercito Libertador 441, Santiago, Chile}

\author[0000-0002-8686-8737]{Franz E. Bauer}
\affiliation{Instituto de Astrof{\'{i}}sica, Facultad de F{\'{i}}sica, Pontificia Universidad Cat{\'{o}}lica de Chile, Campus San Joaqu{\'{i}}n, Av. Vicu\~{n}a Mackenna 4860, Macul Santiago, Chile, 7820436} 
\affiliation{Centro de Astroingenier{\'{i}}a, Facultad de F{\'{i}}sica, Pontificia Universidad Cat{\'{o}}lica de Chile, Campus San Joaqu\'{i}n, Av. Vicu\~{n}a Mackenna 4860, Macul Santiago, Chile, 7820436} 
\affiliation{Millennium Institute of Astrophysics, Nuncio Monse{\~{n}}or S{\'{o}}tero Sanz 100, Of 104, Providencia, Santiago, Chile}

\author[0000-0001-9070-9969]{Rich Bielby}
\affiliation{Centre for Extragalactic Astronomy, Department of Physics, Durham University, South Road, Durham, DH1 3LE, UK}

\author[0000-0002-0167-2453]{W. N. Brandt}
\affiliation{Department of Astronomy \& Astrophysics, 525 Davey Lab, The Pennsylvania State University, University Park, PA 16802, USA}
\affiliation{Institute for Gravitation and the Cosmos, The Pennsylvania State University, University Park, PA 16802, USA}
\affiliation{Department of Physics, 104 Davey Laboratory, The Pennsylvania State University, University Park, PA 16802, USA}

\author[0000-0001-9947-6911]{Colin J. Burke}
\affiliation{Department of Astronomy, University of Illinois Urbana-Champaign, Urbana, IL 61801, USA}

\author{Jorge Casares}
\affiliation{Instituto de Astrof\'isica de Canarias, E-38205 La Laguna, Tenerife, Spain}
\affiliation{Departamento de Astrof\'isica, Universidad de La Laguna, E-38206 La Laguna, Tenerife, Spain}

\author[0000-0002-9932-1298]{Yu-Ching Chen}
\affiliation{Department of Astronomy, University of Illinois Urbana-Champaign, Urbana, IL 61801, USA}

\author[0000-0003-3242-7052]{Gisella De Rosa}
\affiliation{Space Telescope Science Institute, 3700 San Martin Dr., Baltimore, MD 21218, USA}

\author[0000-0001-8251-933X]{Alex Drlica-Wagner}
\affiliation{Fermi National Accelerator Laboratory, P.O.\ Box 500, Batavia, IL 60510, USA}
\affiliation{Kavli Institute for Cosmological Physics, University of Chicago, Chicago, IL 60637, USA}
\affiliation{Department of Astronomy and Astrophysics, University of Chicago, Chicago IL 60637, USA}

\author[0000-0002-4459-9233]{Tom Dwelly}
\affiliation{Max-Planck Institute for Extraterrestrial Physics, Giessenbachstrasse 1,  Garching, 85748, Germany}

\author{Alice Eltvedt}
\affiliation{Centre for Extragalactic Astronomy, Department of Physics, Durham University, South Road, Durham, DH1 3LE, UK}

\author[0000-0003-0042-6936]{Gloria Fonseca Alvarez}
\affiliation{NSF’s NOIRLab, 950 N. Cherry Ave., Tucson, AZ 85719, USA}

\author[0000-0002-3767-299X]{Jianyang Fu(傅健洋)}
\affiliation{Department of Astronomy, University of Illinois Urbana-Champaign, Urbana, IL 61801, USA}

\author[0000-0002-5211-0020]{Cesar Fuentes}
\affiliation{Departamento de Astronom\'{i}a, Universidad de Chile, Camino del Observatorio 1515, Las Condes, Santiago, Chile.}
\affiliation{Centro de Excelencia en Astrof\'{i}sica y Tecnolog\'{i}as Afines (CATA), Chile}
\affiliation{Millennium Institute of Astrophysics (MAS), Chile}

\author[0000-0002-9154-3136]{Melissa L. Graham}
\affiliation{DIRAC Institute, Department of Astronomy, University of Washington, 3910 15th Avenue NE, Seattle, WA 98195, USA}

\author[0000-0001-9920-6057]{Catherine~J.~Grier}
\affiliation{Department of Astronomy, University of Wisconsin-Madison, Madison, WI 53706, USA} 
\affiliation{Steward Observatory, The University of Arizona, 933 North Cherry Avenue, Tucson, AZ 85721, USA}

\author[0000-0002-2582-0190]{Nathan Golovich}
\affiliation{Lawrence Livermore National Laboratory, 7000 East Ave, Livermore, CA 94550, USA}

\author[0000-0002-1763-5825]{Patrick B. Hall}
\affiliation{Department of Physics \& Astronomy, York University, 4700 Keele St., Toronto, ON M3J 1P3, Canada} 

\author[0000-0003-0049-5210]{Patrick Hartigan}
\affiliation{Department of Physics and Astronomy, Rice University, 6100 S. Main St., TX 77005-1892, USA}

\author[0000-0003-1728-0304]{Keith Horne}
\affiliation{University of St Andrews, SUPA School of Physics \& Astronomy, North Haugh, St Andrews KY16 9SS, Scotland, UK}

\author[0000-0002-6610-2048]{Anton M. Koekemoer}
\affiliation{Space Telescope Science Institute, 3700 San Martin Dr., Baltimore, MD 21218, USA} 

\author{Mirko Krumpe}
\affiliation{Leibniz Institute for Astrophysics Potsdam (AIP), An der Sternwarte 16, 14482 Potsdam, Germany}

\author[0000-0002-0311-2812]{Jennifer I. Li}
\affiliation{Department of Astronomy, University of Michigan, Ann Arbor, MI, 48109, USA}

\author[0000-0003-1731-0497]{Chris Lidman}
\affiliation{Research School of Astronomy and Astrophysics, Australian National University, Canberra, ACT 2611, Australia}
\affiliation{Centre for Gravitational Astrophysics, College of Science, The Australian National University, Canberra ACT 2601, Australia}

\author[0000-0002-0036-1696]{Umang Malik}
\affiliation{Research School of Astronomy and Astrophysics, Australian National University, Canberra, ACT 2611, Australia}

\author[0000-0003-2385-6904]{Amelia Mangian}
\affiliation{Information Trust Institute, University of Illinois Urbana-Champaign, Urbana, IL 61801, USA}

\author[0000-0002-0761-0130]{Andrea Merloni}
\affiliation{Max-Planck Institute for Extraterrestrial Physics, Giessenbachstrasse 1,  85748, Garching, Germany}

\author[0000-0001-5231-2645]{Claudio Ricci}
\affiliation{N\'ucleo de Astronom\'ia de la Facultad de Ingenier\'ia, Universidad Diego Portales, Av. Ej\'ercito Libertador 441, Santiago, Chile} 
\affiliation{Kavli Institute for Astronomy and Astrophysics, Peking University, Beijing 100871, China}

\author[0000-0001-7116-9303]{Mara Salvato}
\affiliation{Max-Planck Institute for Extraterrestrial Physics, Giessenbachstrasse 1,  85748, Germany}

\author[0000-0003-4877-7866]{Rob Sharp}
\affiliation{Research School of Astronomy and Astrophysics, Australian National University, Canberra, ACT 2611, Australia}

\author[0000-0002-8501-3518]{Zachary Stone}
\affiliation{Department of Astronomy, University of Illinois Urbana-Champaign, Urbana, IL 61801, USA}
\affiliation{Center for AstroPhysical Surveys, National Center for Supercomputing Applications, University of Illinois Urbana-Champaign, Urbana, IL 61801, USA}

\author[0000-0003-4580-3790]{David E. Trilling}
\affiliation{Department of Astronomy and Planetary Science, Northern Arizona University, Flagstaff, AZ 86011, USA}

\author[0000-0002-4283-5159]{Brad E. Tucker}
\affiliation{Research School of Astronomy and Astrophysics, Australian National University, Canberra, ACT 2611, Australia}
\affiliation{National Centre for the Public Awareness of Science, Australian National University, Canberra, ACT 2601, Australia}
\affiliation{The Australian Research Council Centre of Excellence for All-Sky Astrophysics in 3 Dimension (ASTRO 3D), Australia}

\author[0000-0003-2812-8607]{Di Wen}
\affiliation{Department of Astronomy, University of Illinois Urbana-Champaign, Urbana, IL 61801, USA}

\author[0000-0001-8171-5507]{Zachary Wideman}
\affiliation{Department of Physics and Astronomy, Texas A\&M University, 4242 TAMU, College Station, TX 77843, USA}

\author[0000-0002-1935-8104]{Yongquan Xue}
\affiliation{CAS Key Laboratory for Research in Galaxies and Cosmology, Department of Astronomy, University of Science and Technology of China, Hefei 230026, China}

\author[0000-0003-0644-9282]{Zhefu Yu}
\affiliation{Kavli Institute for Particle Astrophysics and Cosmology, Stanford University, 452 Lomita Mall, Stanford, CA 94305, USA}

\author[0000-0002-2250-730X]{Catherine Zucker}
\affiliation{Center for Astrophysics $\vert$ Harvard \& Smithsonian, 60 Garden Street, Cambridge, MA 02138, USA}

%\author{others}

%\author{program PIs}

%\author{Observers}

%% Note that the \and command from previous versions of AASTeX is now
%% depreciated in this version as it is no longer necessary. AASTeX 
%% automatically takes care of all commas and "and"s between authors names.

%% AASTeX 6.31 has the new \collaboration and \nocollaboration commands to
%% provide the collaboration status of a group of authors. These commands 
%% can be used either before or after the list of corresponding authors. The
%% argument for \collaboration is the collaboration identifier. Authors are
%% encouraged to surround collaboration identifiers with ()s. The 
%% \nocollaboration command takes no argument and exists to indicate that
%% the nearby authors are not part of surrounding collaborations.

%% Mark off the abstract in the ``abstract'' environment. 
\begin{abstract}
High-quality Extragalactic Legacy-field Monitoring (HELM) is a long-term observing program that photometrically monitors several well-studied extragalactic legacy fields with the Dark Energy Camera (DECam) imager on the CTIO 4m Blanco telescope. Since Feb 2019, HELM has been monitoring regions within COSMOS, XMM-LSS, CDF-S, S-CVZ, ELAIS-S1, and SDSS Stripe 82 with few-day cadences in the $(u)gri(z)$ bands, over a collective sky area of  $\sim 38$\,deg${\rm ^2}$. The main science goal of HELM is to provide high-quality optical light curves for a large sample of active galactic nuclei (AGNs), and to build decades-long time baselines when combining past and future optical light curves in these legacy fields. These optical images and light curves will facilitate the measurements of AGN reverberation mapping lags, as well as studies of AGN variability and its dependences on accretion properties. In addition, the time-resolved and coadded DECam photometry will enable a broad range of science applications from galaxy evolution to time-domain science. We describe the design and implementation of the program and present the first data release that includes source catalogs and the first $\sim 3.5$~years of light curves during 2019A--2022A. 
\end{abstract}

%% Keywords should appear after the \end{abstract} command. 
%% The AAS Journals now uses Unified Astronomy Thesaurus concepts:
%% https://astrothesaurus.org
%% You will be asked to selected these concepts during the submission process
%% but this old "keyword" functionality is maintained in case authors want
%% to include these concepts in their preprints.

\section{Introduction}\label{sec:intro}

%\red{Mingyang please add author info from the google doc. The list of authors is here in the comments:} %https://docs.google.com/document/d/1RzrN1QbMKTVOTbf_yJvuR7nBeXFdvGrLEcpcf0csL7U/edit}\blue{Done. Authors please check again the author name, ORCID, and affliation are correct.}

Large-scale optical time-domain imaging surveys such as Sloan Digital Sky Survey  \citep[SDSS;][]{SDSS} supernova survey \citep{SDSS_supernova}, Pan-STARRS1 Medium Deep Survey \citep{PS1, PS1_MDS}, Zwicky Transient Facility \citep[ZTF;][]{ZTF}, and Wide Field Survey Telescope \citep{WFST}
%\citep[e.g.,][]{SDSS, PS1, DES-overview, WFST,Aleo_2023} 
are rapidly shaping our understanding of the dynamic and variable universe. In addition to transients and explosive events from stellar systems, these time-domain surveys provide extremely valuable information to probe the physical mechanisms that drive the persistent variability from variable stars and active galactic nuclei (AGNs). Variability studies of AGNs also enable the measurements of their black hole masses via the reverberation mapping (RM) technique that measures the time lag between the driving continuum variability and the delayed response in the broad-line emission \citep[e.g.,][]{Blandford_McKee_1982,Peterson_1993}. In particular, high-quality photometric light curves with long time baselines and adequate cadences are necessary to secure a RM lag measurement and periodicity searches for binary supermassive black hole candidates when combined with dedicated spectroscopic monitoring data \cite[e.g.,][]{Shen_etal_2015a, search_binary_BH}.

The SDSS-V Black Hole Mapper (BHM) program \citep{Kollmeier_etal_2017} is conducting a multi-year (2020--2027), multi-object spectroscopic RM campaign (BHM-RM) in several legacy extragalactic fields. In order to provide the required photometric light curves that sample the AGN continuum variability, we are conducting a long-term photometric monitoring campaign with the Dark Energy Camera \citep[DECam;][]{DECam} on the 4m CTIO Blanco telescope. The fields chosen by the SDSS-V BHM-RM program coincide with several well-studied extragalactic fields, each with ample multi-wavelength data and/or earlier and future photometric light curves therein \citep[e.g.,][]{Lacy+2021MNRAS, Zou+2022ApJS}. For example, most of these fields coincide with deep monitoring fields in the Pan-STARRS1 survey \citep{PS1} and the Dark Energy Survey supernovae program \citep{Abbott+2021ApJS}. The combined optical light curves for most of these monitoring fields will span more than 25 years once combined with 10-yr light curves from the Legacy Survey of Space and Time (LSST) with the Vera C. Rubin Observatory. Therefore these DECam monitoring data will provide legacy value for broad science in these fields. 

In this paper we provide an overview of the DECam monitoring program and its current status, and present the first data release\footnote{\url{https://ariel.astro.illinois.edu/helm/}} from the first $\sim 3.5$ years of observations (2019-2022). In Section~\ref{sec:data} we provide a technical overview of the High-quality Extragalactic Legacy-field Monitoring (HELM) program. In Section~\ref{sec:application} we describe the data products, with examples to showcase the science applications. We summarize and provide an outlook in Section~\ref{sec:con}. We adopt AB magnitude system throughout the paper.

\begin{deluxetable*}{lllllcc}[ht]
\tablenum{1}
\caption{HELM observation summary \label{table1}}
\tablehead{
\colhead{Field} & \colhead{Pointing} & \colhead{Pointing center} & \colhead{ProgID} & \colhead{Filter} & \colhead{Nexp} & \colhead{$5\sigma$ depth (mag)}\\
\colhead{(1)} & \colhead{(2)} & \colhead{(3)} & \colhead{(4)} & \colhead{(5)} & \colhead{(6)} & \colhead{(7)}
}
\startdata
CDF-S & C1 & 03h37m06s $-$27d06m43s & C & $griz$ & 54-36-36-18 & 22.9-23.3-23.1-22.6\\
 & C2 & 03h37m06s $-$29d05m18s & A-C & $griz$ & 54-36-36-18\\
 & C3 & 03h30m36s $-$28d06m01s & A-C & $griz$ & 248-190-187-34\\
COSMOS & CO1 & 10h00m00s $+$03d05m56s & A-C-E-F-G & $gri$ & 362-295-295 & 22.8-23.1-22.9\\
 & CO2 & 09h56m53s $+$01d44m58s & A-E-F-G & $gri$ & 278-232-235\\
 & CO3 & 10h03m07s $+$01d44m59s & A-E-F-G & $gri$ & 267-226-232\\
ELAIS-S1 & E1 & 00h31m30s $-$43d00m35s & A-C-E-F-G& $griz$ & 387-340-349-28 & 22.8-23.2-23.1-22.5\\
& E2 & 00h38m00s $-$43d59m53s & A-C-E-F-G & $griz$ & 365-332-322-302\\
S-CVZ & S-CVZ & 06h00m00s $-$66d33m39s& A-C & $ugriz$ & 33-223-196-196-24 & 21.1-22.8-22.9-22.7-22.0\\
Stripe 82 & S1 & 02h51m17s $-$00d00m03s & A-B-C-H & $griz$ & 87-50-47-16 & 22.7-23.5-23.4-22.8\\
& S2 & 02h44m47s $-$00d59m20s & A-B-C-H & $griz$ & 82-47-47-17\\
XMM-LSS & X1 & 02h17m54s $-$04d55m48s & A-C & $griz$ & 127-107-102-20 & 22.6-22.9-22.7-22.4\\
 & X2 & 02h22m40s $-$06d24m46s & A-C & $griz$ & 127-107-102-20\\
 & X3 & 02h25m48s $-$04d36m04s & A-C & $griz$ & 127-107-102-20\\
\enddata
\tablecomments{Col. (1) Field name. Col. (2) Pointing name. Col. (3) Median coordinate of pointing center. Typical pointing accuracy of each exposure is $\sim$10\arcsec. Col. (4) Program ID. A: 2019A-0065; B: 2019B-0219; C: 2019B-0910; D: 2021A-0037; E: 2021A-0113; F: 2021B-0149; G: 2022A-724693; H: 2022B-175073. Col. (5) Filter used. Col. (6) Number of successfully reduced exposures in each band, separated by dashes. Most of the nightly epochs in the HELM program consist of two back-to-back exposures per pointing per filter. Col. (7) Median $5\sigma$ depth of individual exposures for all pointings of a given field estimated from $m_{\rm PSF}$.}
\end{deluxetable*}

\section{Observations and Data Processing}\label{sec:data}

\begin{figure*}[t]
\centering
\includegraphics[width=\textwidth]{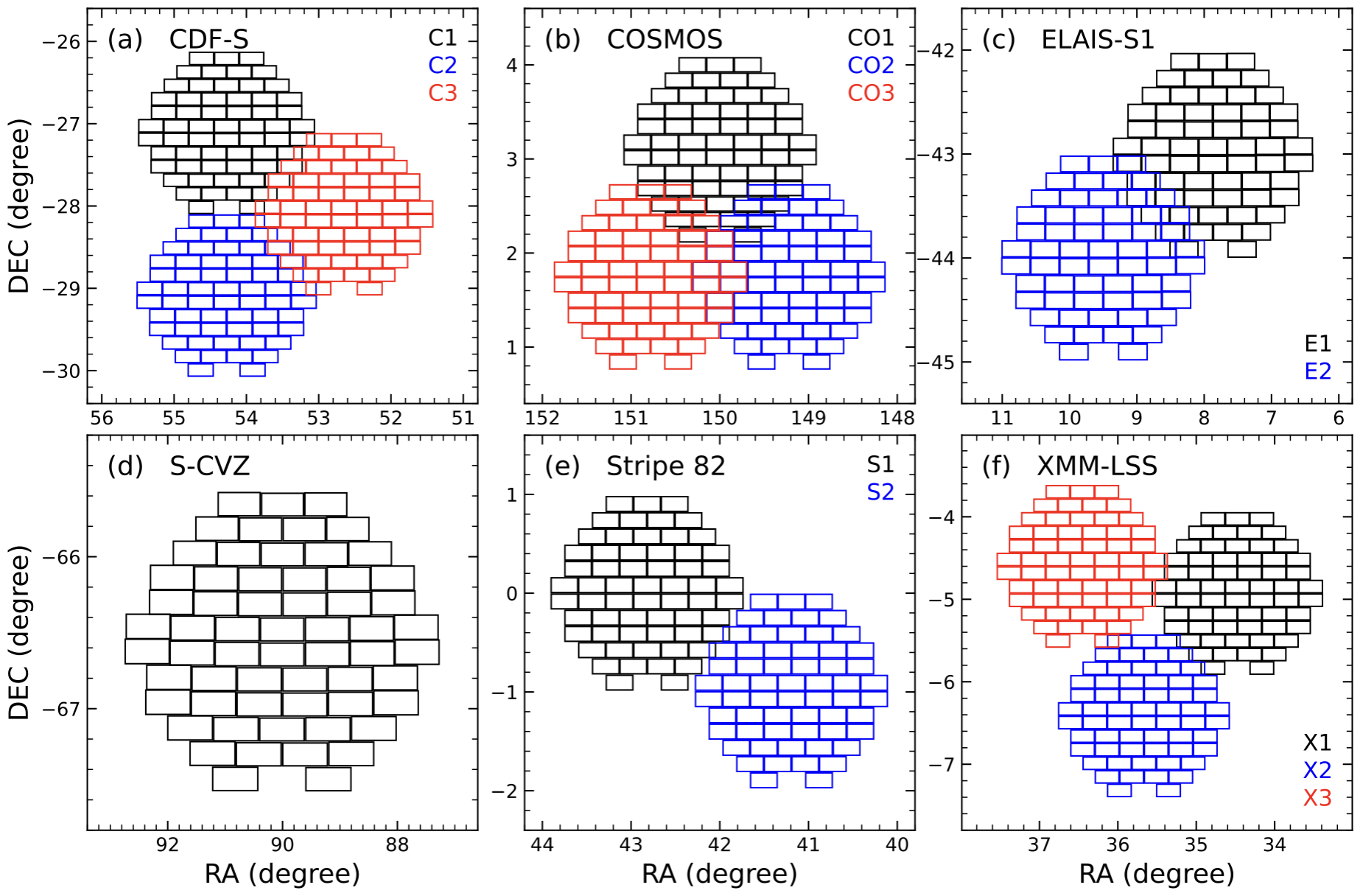}
\caption{The layout of HELM pointings in (a) CDF-S, (b) COSMOS, (c) ELAIS-S1, (d) S-CVZ, (e) SDSS Stripe 82, and (f) XMM-LSS fields. Pointings 1, 2, and 3 are indicated in black, blue, and red colors, respectively.}
\label{fig1}
\end{figure*}

HELM started observations in the 2019A semester (Feb-Jul, 2019) and is currently ongoing. During the early phase of the program (2019A-2020B), a cadence of $\sim 6$ days was adopted. Since 2021A, the cadence was increased to $\sim 3$~days and a new field (South-Continuous Viewing Zone; S-CVZ) was added to the field list. For each epoch, a set of consecutive exposures in the $gri$ bands are taken, with 2 back-to-back exposures for each band. The nominal per-epoch exposure times are 160~s, 140~s, and 180~s in the three bands, respectively. A total of 14 DECam pointings (each with a $2.7\,{\rm deg^2}$ field-of-view) received regular monitoring, most of which are for the BHM-RM program. However, some pointings are included in different programs (PIs: X. Liu, P. Martini) that have slightly different science goals, including periodicity searches of binary supermassive black hole candidates and RM lag measurement. We consolidated these different DECam monitoring programs with AGN science focuses and processed these data with the same pipeline. Table~\ref{table1} summarizes the details of the monitoring fields. {Three fields (CDF-S, XMM-LSS, and COSMOS) contain three pointings, while two fields (ELAIS-S1 and SDSS Stripe 82) contain two pointings. The layouts of the HELM fields are shown in Figure~\ref{fig1}.} These particular pointings are aligned with past monitoring within the DES supernovae program \citep[e.g.,][]{DES_supernova1, DES_supernova2}. 

The first 2 semesters of HELM (2019A-2019B) were carried out on shared nights with other non-monitoring DECam programs. Since 2020A, HELM has been part of a collection of regular DECam monitoring programs that pool observing time and observers. In a few subsequent semesters, to improve the overall observing efficiency, some pointings of HELM were observed under the public DECam monitoring program that also targeted the COSMOS and ELAIS-S1 fields \citep{Graham_etal_2023}, but to greater depths than those of the HELM program. These public DECam data are included in our data reduction and data products. The monitoring program suffered significant time loss in 2020A due to the COVID-19 pandemic, and substantial loss in the second half of 2020B due to DECam instrumental problems. 

\begin{figure*}[t]
\centering
\includegraphics[width=\textwidth]{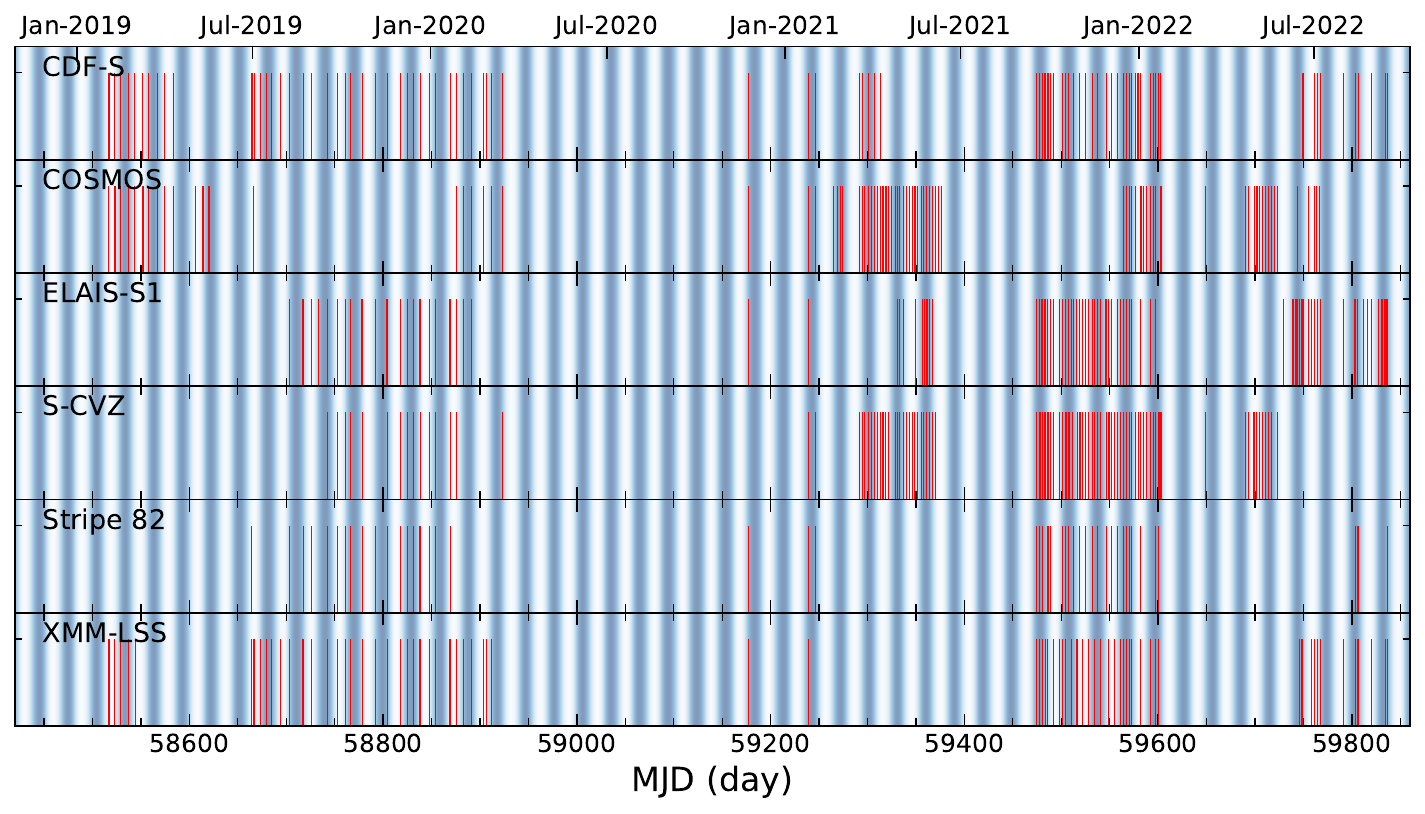}
\caption{Cadence of the HELM in different fields, with vertical red lines indicating individual exposures. Moon illumination (illuminated fraction of the moon) is indicated by the blue gradient (white$=$no moonlight; dark blue$=$maximum moonlight) in the background. }
\label{fig2}
\end{figure*}

At the beginning of the program during 2019A, we observed some fields with additional filters to provide more wavelength coverage. The S-CVZ field was observed in the $ugriz$ filters. For these additional imaging, $griz$ has 61 available CCD chips (except CCD \#61), while $u$ has 60 chips with an additional CCD \#2 not available. Table~\ref{table1} summarizes the HELM observations including fields, pointings, observing program IDs, filters, and numbers of exposures up to 13 September 2022. Figure~\ref{fig2} illustrates the cadence of each field.

%\red{Table 1 will summarize the fields, observing progID and status -- make sure we include public programs that cover our fields, e.g., COSMOS. }

\subsection{Data Reduction}

The raw HELM observations were processed using the Dark Energy Survey (DES) ``Final Cut'' pipeline \citep[see][for an overview]{Morganson_etal_2018} with the same configuration of \texttt{Sextractor} \citep{1996A&AS..117..393B} as that used in the production campaigns for the second data release (DR2) of DES \citep[DES-DR2;][]{Abbott+2021ApJS} and DECam Local Volume Exploration-DR2 \citep{Drlica-Wagner_etal_2022}. During the years that the DES Collaboration was actively taking data, the calibration products (e.g., bias, flat, sky-subtraction templates, secondary flat-field or ``star-flats'', and the relative astrometric offsets among the CCDs) were monitored and updated as needed as part of that program.  Processing of observations taken after the DES observations ended in January 2019 have, thus far, relied on the DES legacy calibration products from that period.  In Section \ref{chk_cal} we assess the overall systematic astrometric and photometric performance to understand the limitations this might have on the temporal monitoring by HELM.  
%\red{Blame Robert... I think in order to follow through on the above... means you should consider making plots showing overall performance vs time.  That is what you ultimately need/want isn't it?} \red{Monika, Doug and Robert will add the DECADE data reduction description. } Use \texttt{Sextractor} to perform source detection, segmentation, and measurement.

\subsubsection{Magnitude Zeropoint}\label{sec:zp}
A magnitude zeropoint (zp) pipeline is run separately for all of the CCD images. We use the \texttt{expCalib} (S. Allam \& R. Gruendl, private communication) code to match objects in the Final Cut catalogs to the APASS 9 catalog \citep{APASSDR9}. A $1\arcsec$-radius match was performed for each CCD in one exposure. If the number of matched objects is low or a match was not found, the average zp value among all CCDs is adopted. The expCalib will not return a match (or will return a low number of matches) in two cases:  - APASS catalog has no coverage in the requested region of the sky, or the magnitude range of stars in the APASS catalog does not overlap with the brightness of objects detected in the CCD  (APASS sources are brighter and appear saturated in the DECam data). Among 7816 successfully reduced exposures (corresponding to 476743 CCD individual images), magnitude zero points are available from the expCalib for all but 124 exposures (2, 49, 28, 44, and 1 in the $ugriz$ bands, respectively) and 7562 CCD images ($1.6\%$). 

Using four fields associated with 10 pointings of our HELM DECam monitoring program (CDF-S, ELAIS-S1, SDSS Stripe 82, and XMM-LSS) that are covered by the footprint of 6 years of DES science operations of $\sim5000$ deg$^2$ from DES-DR2 \citep{Abbott+2021ApJS}, we derive independent magnitude zeropoints for all exposures by cross-matching sources detected in our observations with those from the DES-DR2 source catalog. The DES-DR2 catalog is constructed using coadded images, which has typically $\sim10$ times longer exposure time than our single-exposure images. Based on the extent of the source, we derive two versions of magnitude zeropoints, one based on point sources using $m_{\rm PSF}$ and another based on all the sources using $m_{\rm AUTO}$ to increase the number of available objects. We first select all the reliably extracted (\texttt{IMAFLAGS\_ISO=0},  \texttt{FLAGS<4}), high signal-to-noise ratio (SNR; $m_{\rm PSF}$ error $<0.1$ mag, which corresponds to an SNR of 10) sources from our observations. We use a relatively loose criterion of \texttt{CLASS\_STAR $>0.8$} to select point-like sources, which ensures we do not miss a significant number of point sources and avoids including many extended sources. For the DES-DR2 catalog, we apply the same selection criteria of object quality (\texttt{IMAFLAGS\_ISO=0},  \texttt{FLAGS<4} in $i$ band) and select point-like sources using \texttt{$0 \leq$ EXTENDED\_COADD $\leq 1$} following \citet{Abbott+2021ApJS}. 

\begin{figure}[t]
\centering
\includegraphics[width=0.49\textwidth]{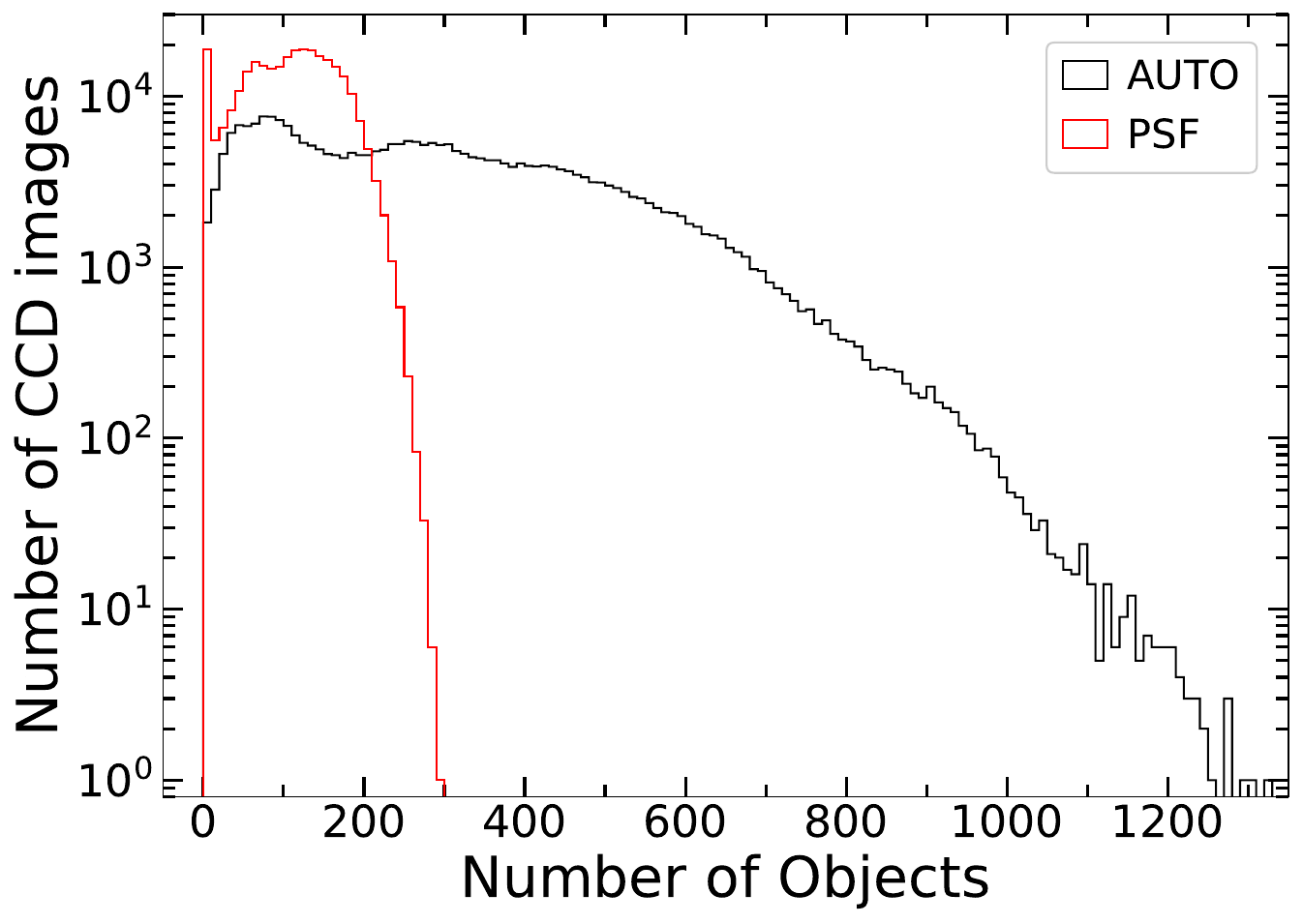}
\caption{Histograms of objects used to calibrate the magnitude zeropoint for each CCD image of every exposure. Histogram for point sources using $m_{\rm PSF}$ is in red (PSF), while histogram for all sources using $m_{\rm AUTO}$ is in black.}
\label{fig3}
\end{figure}

\begin{figure}[t]
\centering
\includegraphics[width=0.49\textwidth]{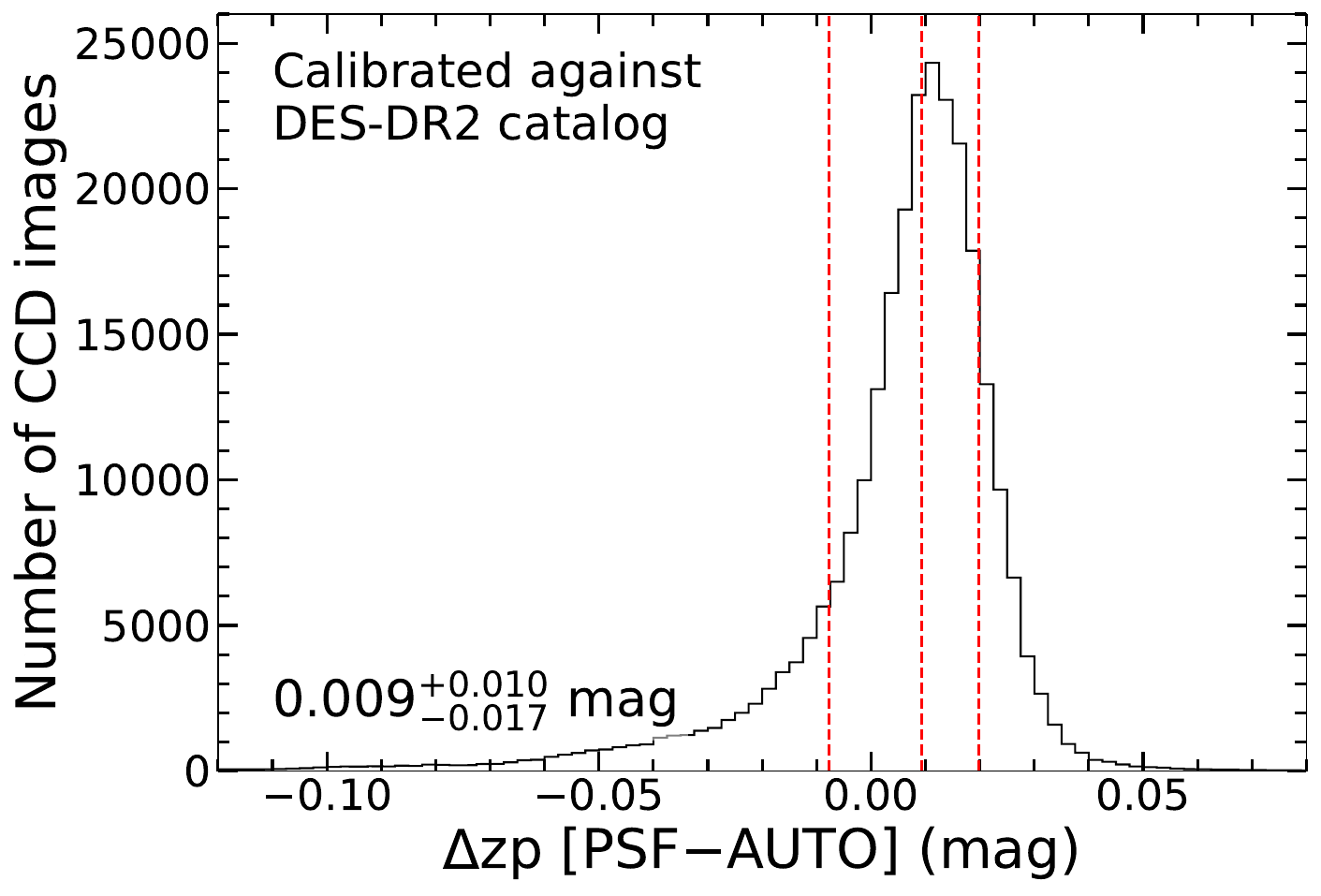}
\caption{Comparison of magnitude zeropoints (zp) calibrated against the DES-DR2 catalog using point sources (PSF) and all sources (AUTO). Only CCD images with at least 10 sources used for calibration are included. Red vertical dashed lines indicate 16th, 50th, and 84th percentiles of the distribution, with statistics shown in the lower-left corner.}
\label{fig4}
\end{figure}

The magnitude zp of each CCD chip is derived using the 3-$\sigma$ clipped median difference between magnitudes of objects from our single-exposure catalog and those from DES-DR2 catalog, with associated error equal to the standard deviation divided by the root of the number of the remaining objects. Figure~\ref{fig3} compares the number of objects used to derive the magnitude zps. Magnitude zps derived from all sources make use of on average 2.7 times more objects compared to those derived from point sources only (295 versus 108). Limiting to CCD images with at least 10 applicable objects for zp calibration, we find a small difference of 0.009 mag with 16th and 84th percentiles of the distribution of $-$0.008 and 0.019 mag between the two sets of zp, respectively (Figure~\ref{fig4}). We find that the difference between two zps is correlated with seeing, with larger $m_{\rm PSF}$-based zp in exposures with better seeing. This may be related to slight difference in aperture loss at different seeing conditions. We also compare the zp from the DECam pipeline with that from DES-DR2 using all sources (Table~\ref{table2}). The median differences are $<0.01$ mag (slightly larger with 0.014 mag in the E2 field) with a scatter of $\la 0.04$ mag. The final zp of a CCD image is determined with the priority decreasing from DES-DR2 AUTO magnitude to DES-DR2 PSF magnitude to DECam pipeline. We only adopt DES-DR2-based zp if the number of objects used to calibrate zp is greater or equal to 10. Finally, we successfully obtain zp for 99.9\% of the CCD images, with DES-DR2 AUTO contributing to the vast majority of them (99.4\%).

\begin{deluxetable*}{cRRRRRRRRR}[ht]
\tablenum{2}
\caption{Comparison of HELM zp, $m_{\rm PSF}$, and astrometry to those from DES-DR2 \label{table2}}
\tablehead{
\colhead{Pointing} & \colhead{$\Delta$zp} & \colhead{$\Delta m_{\rm PSF}$ [$g$]} & \colhead{$\Delta m_{\rm PSF}$ [$r$]} & \colhead{$\Delta m_{\rm PSF}$ [$i$]} & \colhead{$\Delta m_{\rm PSF}$ [$z$]} & \colhead{$\Delta d$ [$g$]} & \colhead{$\Delta d$ [$r$]} & \colhead{$\Delta d$ [$i$]} & \colhead{$\Delta d$ [$z$]}\\
\nocolhead{Field} & \colhead{mag} &  \colhead{(mag)} & \colhead{(mag)} & \colhead{(mag)} & \colhead{(mag)} & \colhead{(\arcsec)} & \colhead{(\arcsec)} & \colhead{(\arcsec)} & \colhead{(\arcsec)}
}
\startdata
C1 & -0.007^{+0.017}_{-0.032} & -0.019_{-0.042}^{+0.038} & -0.026_{-0.042}^{+0.036}& -0.011_{-0.029}^{+0.026} & -0.015_{-0.029}^{+0.024} & 0.059_{-0.032}^{+0.054} & 0.052_{-0.028}^{+0.050} & 0.050_{-0.027}^{+0.047} & 0.050_{-0.027}^{+0.045} \\
C2 & -0.006^{+0.015}_{-0.024}& -0.015_{-0.036}^{+0.035} & -0.020_{-0.039}^{+0.034} & -0.013_{-0.028}^{+0.031} & -0.015_{-0.028}^{+0.028} & 0.059_{-0.031}^{+0.054} & 0.052_{-0.028}^{+0.050} & 0.051_{-0.027}^{+0.047} & 0.051_{-0.027}^{+0.046} \\
C3 & -0.005^{+0.018}_{-0.031} & -0.016_{-0.041}^{+0.036} & -0.018_{-0.041}^{+0.032} & -0.012_{-0.030}^{+0.028} & -0.018_{-0.030}^{+0.025} & 0.066_{-0.036}^{+0.063} & 0.060_{-0.033}^{+0.060} & 0.058_{-0.032}^{+0.057} & 0.052_{-0.028}^{+0.048} \\
E1 & 0.004^{+0.029}_{-0.043} & -0.029_{-0.042}^{+0.037} & 0.020_{-0.042}^{+0.039} & 0.007_{-0.033}^{+0.032} & -0.004_{-0.028}^{+0.028} & 0.075_{-0.040}^{+0.072} & 0.068_{-0.036}^{+0.066} & 0.065_{-0.035}^{+0.061} & 0.053_{-0.028}^{+0.048} \\
E2 & -0.014^{+0.024}_{-0.035} & -0.037_{-0.039}^{+0.036} & -0.005_{-0.034}^{+0.040} & -0.014_{-0.034}^{+0.032} & -0.011_{-0.025}^{+0.026} & 0.073_{-0.039}^{+0.069} & 0.067_{-0.036}^{+0.065} & 0.064_{-0.034}^{+0.061} & 0.054_{-0.029}^{+0.048} \\
S1 & -0.008^{+0.044}_{-0.029} & -0.023_{-0.041}^{+0.045} & -0.009_{-0.032}^{+0.047} & 0.028_{-0.043}^{+0.051} & 0.009_{-0.026}^{+0.037} & 0.066_{-0.036}^{+0.063} & 0.058_{-0.032}^{+0.061} & 0.057_{-0.031}^{+0.058} & 0.050_{-0.027}^{+0.047} \\
S2 & 0.000^{+0.026}_{-0.024} & 0.009_{-0.053}^{+0.052} & -0.010_{-0.033}^{+0.038} & -0.009_{-0.044}^{+0.032} & -0.013_{-0.037}^{+0.030} & 0.066_{-0.036}^{+0.064} & 0.058_{-0.032}^{+0.062} & 0.058_{-0.032}^{+0.059} & 0.052_{-0.028}^{+0.049} \\
X1 & -0.002^{+0.020}_{-0.035} & -0.008_{-0.048}^{+0.041} & -0.015_{-0.045}^{+0.039} & -0.008_{-0.035}^{+0.031} & -0.012_{-0.034}^{+0.030} & 0.068_{-0.036}^{+0.066} & 0.062_{-0.034}^{+0.064} & 0.061_{-0.033}^{+0.059} & 0.055_{-0.030}^{+0.051} \\
X2 & 0.004^{+0.019}_{-0.033} & -0.005_{-0.051}^{+0.044} & -0.008_{-0.044}^{+0.041} & -0.009_{-0.037}^{+0.032} & -0.015_{-0.035}^{+0.029} & 0.070_{-0.038}^{+0.068} & 0.064_{-0.035}^{+0.064} & 0.062_{-0.034}^{+0.061} & 0.056_{-0.030}^{+0.051} \\
X3 & 0.003^{+0.019}_{-0.030} & -0.009_{-0.047}^{+0.042} & -0.006_{-0.041}^{+0.038} & -0.003_{-0.033}^{+0.031} & -0.011_{-0.029}^{+0.028} & 0.070_{-0.038}^{+0.066} & 0.062_{-0.034}^{+0.063} & 0.060_{-0.033}^{+0.059} & 0.053_{-0.028}^{+0.049}\\
\enddata
\tablecomments{ $\Delta \equiv $ our DECam data $-$ DES-DR2. Statistics are 50th percentile with upper (84th $-$ 50th) and lower (16th $-$ 50th) 1$\sigma$ of the distribution as the superscript and subscript, respectively.$\Delta d$ represents the angular distance difference between coordinates.}
\end{deluxetable*}

\begin{figure}[t]
\centering
\includegraphics[width=0.49\textwidth]{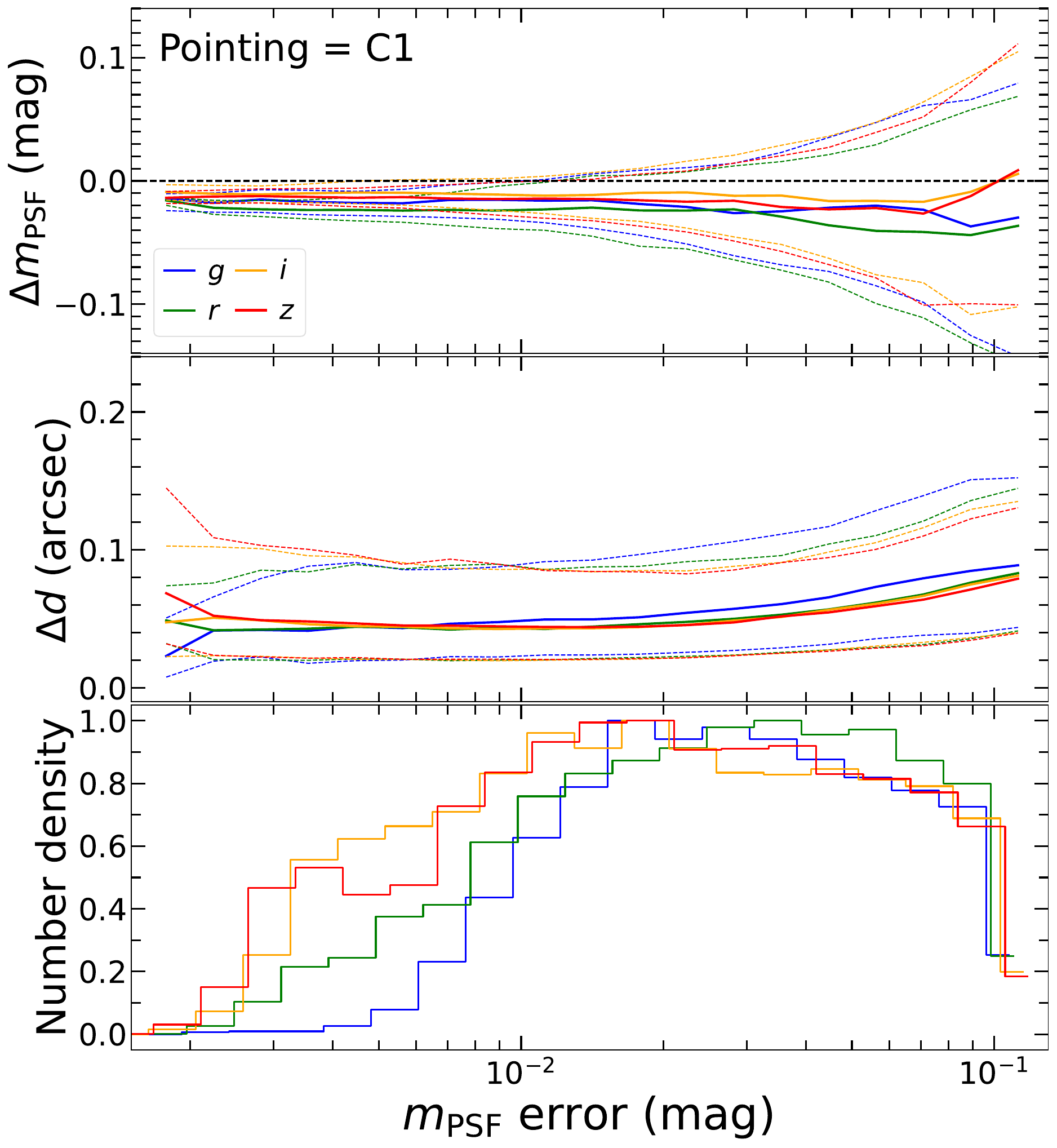}
\caption{Differences in $m_{\rm PSF}$ (top) and coordinates (middle) between our measurements of all observations and those from the DES-DR2 catalog ($\Delta\equiv$ HELM $-$ DES-DR2) as a function of $m_{\rm PSF}$ error for all high SNR ($m_{\rm PSF}$ error $<0.1$ mag) point sources in pointing C1. Normalized source number density ($N/N_{\rm max}$) of $m_{\rm PSF}$ error is shown in the bottom panel. $m_{\rm PSF}$ error is the propagated error by combining our measurement error and DES-DR2 measurement error in quadrature sum. Median values in the $griz$ filters are shown as blue, green, orange, and red solid curves, respectively. Dashed curves in top and middle panels indicate values at 84th and 16th percentiles. Statistics of all 10 pointings are shown in Table~\ref{table2}.} 
\label{fig5}
\end{figure}

\subsection{Photometry and Astrometry Cross-Calibration\label{chk_cal}}
Following the same recipe as in Section~\ref{sec:zp}, we check the accuracy of the HELM pipeline photometry and astrometry independently for the fields that are not covered by the official DES data products (COSMOS and S-CVZ). The DES-DR2 catalog is based on coadded images, with a photometric accuracy of $\sim0.01$ mag and astrometric precision of $\sim 0\farcs03$. We select all the point sources detected in our observations and crossmatch with the DES-DR2 catalog for all the reliably extracted (\texttt{IMAFLAGS\_ISO=0},  \texttt{FLAGS<4}), high SNR ($m_{\rm PSF}$ error $<0.1$ mag) sources. We select point sources from the DES-DR2 catalog using \texttt{$0 \leq$ EXTENDED\_COADD $\leq 1$}. For our observations, we use a relatively loose criterion of \texttt{CLASS\_STAR $>0.8$}, which ensures we do not miss a significant number of point sources and avoids including many extended sources. 

Figure \ref{fig5} presents an example of comparison of $m_{\rm PSF}$ and coordinate difference ($\Delta d$) as a function of $m_{\rm PSF}$ error in pointing C1. The statistics for all 10 pointings are presented in Table~\ref{table3}. Although various degrees of differences are found in different fields, the source magnitude (median $\Delta m_{\rm PSF} \approx 0.02$ mag) and astrometry ($\Delta d\approx0\farcs06$) of our HELM pipeline are in good agreement with DES-DR2 for high SNR sources. The overall good performances in zp, photometry, and astrometry lend support to the reliability of the HELM pipeline for use in fields not covered by the footprint of DES-DR2 (COSMOS and S-CVZ).

\begin{figure}[t]
\centering
\includegraphics[width=0.49\textwidth]{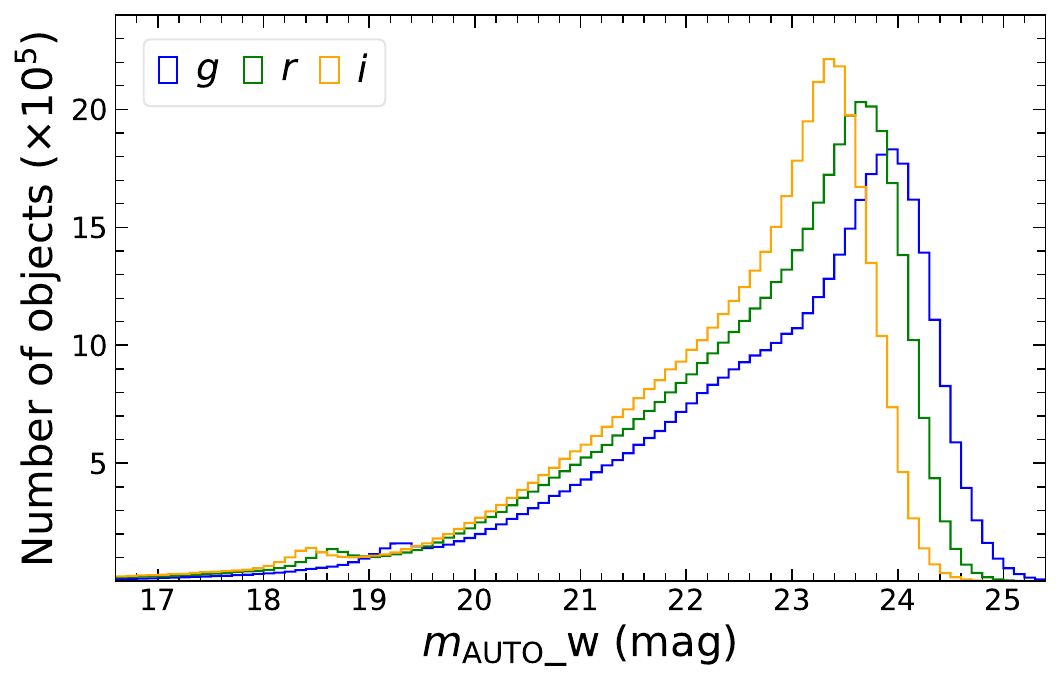}
\caption{Histograms of inverse variance-weighted mean $m_{\rm AUTO}$ in all detected exposures ($m_{\rm AUTO}$\_w; see Equation~\ref{eq1}) in $g$ (blue), $r$ (green), and $i$ (orange) bands for objects in all 6 fields with detection in at least 5 individual exposures.} 
\label{fig6}
\end{figure}

\begin{figure*}[ht]
\centering
\includegraphics[width=\textwidth]{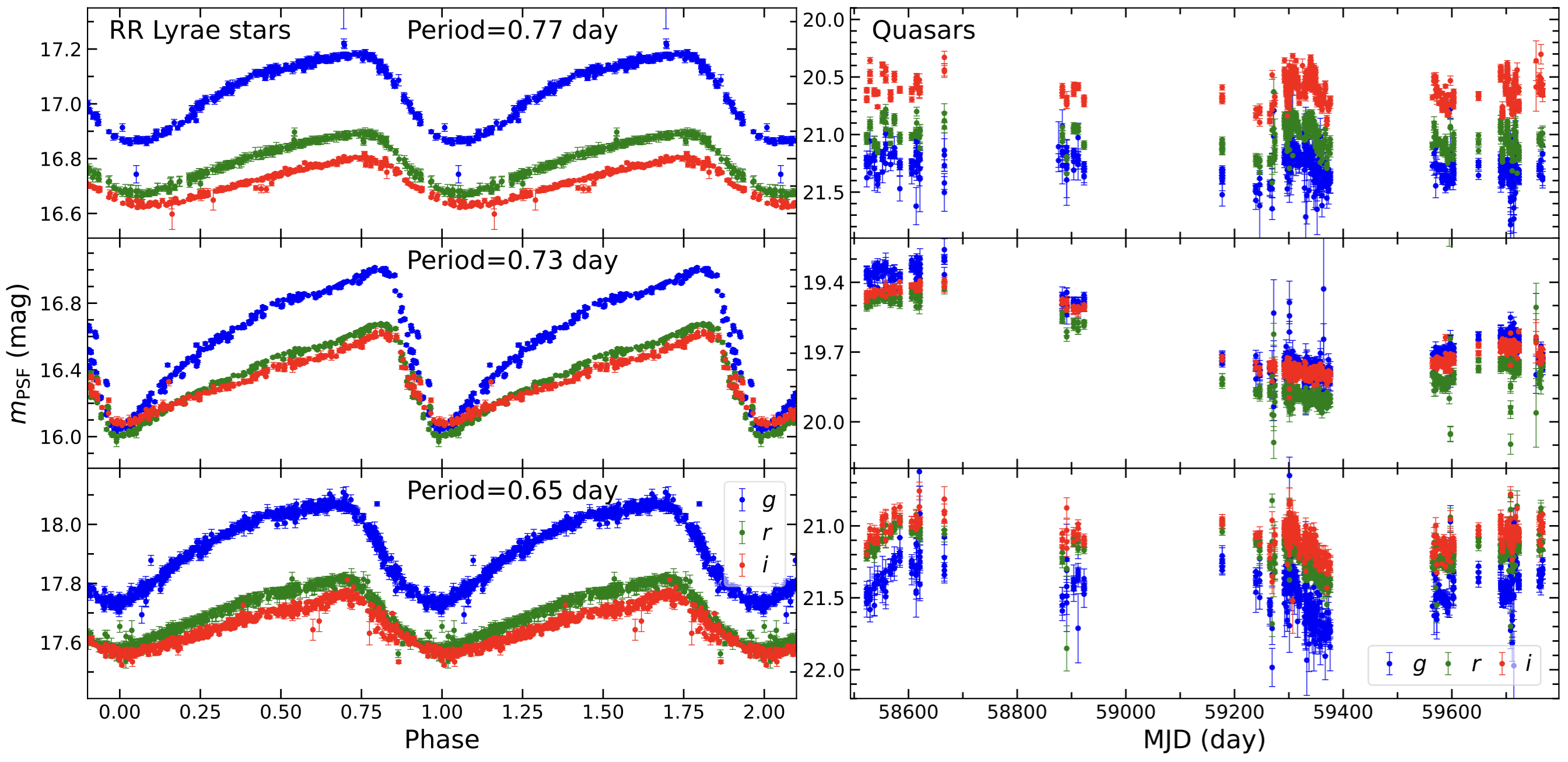}
\caption{Example light curves of RR Lyrae stars (left; in folded phase) and AGNs (right) in the COSMOS field. Blue, green, and red dots represent the $g$, $r$, and $i$ bands, respectively, with error bar indicating the certainty. The periods of RR Lyrae stars are shown at the top of the left panels.}
\label{fig7}
\end{figure*}

\section{Data Products}\label{sec:application}

For this first HELM data release, we include the source catalog covering all compiled fields, as well as the light curves for these sources from the first 3.5 years (2019A-2022A). Our source catalog is built by cross-matching detected sources in our individual images with the DES-DR2 \citep{Abbott+2021ApJS} source catalog for the CDF-S, XMM-LSS, ELAIS-S1, and SDSS Stripe 82 fields, and with the source catalog from the third data release \citep{SSP-DR3} of the Hyper Suprime-Cam (HSC) Subaru Strategic Program \citep[SSP;][]{HSC-SSP} for the COSMOS field. In total, there are 5,404,721, 4,714,784, and 4,059,093 sources with detections in at least 3, 5, and 10 exposures in any band, respectively. Figure~\ref{fig6} shows histograms of inverse variance-weighted mean $m_{\rm AUTO}$ ($m_{\rm AUTO}$\_w; see Section~\ref{sec3.1} for definition) in the $gri$ bands for objects in all 6 fields with detections in at least 5 individual exposures. The peak $m_{\rm AUTO}$\_w is $\sim24.0$, 23.6, and 23.4 mag for the $gri$ bands, respectively, which are deeper compared to those of typical single exposures (Table~\ref{table1}) and representing the best seeing condition and longest exposure time. In future data releases, we will provide individual images, deeper photometric catalogs built from the coadded images, and extended light curves (see Section~\ref{sec:con}). Below we describe in detail our source catalog and light curve compilation. 

\subsection{Source catalogs and light curves}\label{sec3.1}

%\red{Describe the steps to construct the source catalogs in each field and the compilation of light curves. }

The majority of our DECam pointings in the COSMOS field are covered by Deep- and UltraDeep-depth images of HSC SSP. For regions outside the Deep and UltraDeep coverage, we use instead the source catalog from Wide-depth images of HSC SSP. We extract the source catalog with $i$ band \texttt{Kron} magnitude brighter than 25.5 mag from the Hyper Suprime-Cam Legacy Archive\footnote{\url{https://hscla.mtk.nao.ac.jp/doc/}} (\texttt{pdr3\_dud\_rev} for the Deep and UltraDeep catalogs and \texttt{pdr3\_wide} for the Wide catalog) using example script \texttt{PDR3 example 1a}. For the Wide catalog, we further apply the criteria presented in \texttt{PDR3 example 6} script to the $i$ band to keep only ``clean'' sources. For all the other fields except S-CVZ (i.e., CDF-S, ELAIS-S1, Stripe 82, and XMM-LSS), we use DES-DR2 as the base source catalog for catalog construction, as it is deeper than individual exposures of our program. A radius of 0\farcs5 is adopted for crossmatching. 

For the S-CVZ field, we construct the source catalog from scratch as no deep optical catalog in this field is currently available. We first randomly pick an individual exposure catalog as the reference catalog and crossmatch other catalogs in the $griz$ bands to it one by one using a matching radius of 0.5\arcsec. We then iteratively build the preliminary master catalog by adding new sources whose separations from existing sources are larger than 0.5\arcsec. The coordinates of the preliminary master catalog obtained after the first run are determined as the median value of those classified as the same source in individual catalogs. We then perform a second run by crossmatching individual exposure catalogs to the preliminary master catalog and obtain a final catalog in the $griz$ bands. Finally, we crossmatch the $u$ band individual exposure catalogs with the $griz$ catalog, as the $u$ band depth is much shallower compared to that in the $griz$ bands. 

Table~\ref{table3} presents the table format of our median source catalog, including source index, coordinate, magnitudes (PSF and AUTO), extraction flags, star-galaxy classifier (spread\_model and class\_star), and number of exposures. For magnitude, we also provide inverse variance weighted mean (mag\_w), defined as  
\begin{equation}\label{eq1}
   {\rm mag\_w} \equiv \frac{\Sigma(m_i/e_i^2)}{\Sigma(1/e_i^2)}, 
\end{equation}
where $m_i$ and $e_i$ represent an individual measurement and its error in the $i$th exposure, respectively. Its corresponding uncertainty is 
\begin{equation}\label{eq2}
 {\rm mag\_w\_err} = \frac{\sqrt{\Sigma(e_i/e_i^2)^2}}{\Sigma(1/e_i^2)}=\frac{1}{\sqrt{\Sigma(1/e_i^2)}},
\end{equation}
after error propagation. For the S-CVZ field, we provide separate source indices and coordinates for $griz$ and $u$ bands, as the $u$ band is much shallower compared to the other bands. Separation between the median coordinate in the $griz$ bands and that in the $u$ band is also provided. Table~\ref{table4} describes the format of the light curve files of individual objects. The light curves of individual sources can be extracted from our online database\footnote{\url{https://ariel.astro.illinois.edu/helm/}} using source indices provided in Table~\ref{table3}. 

%The astrometry of HSC SSP catalog is calibrated using PanSTARRS1 DR2 catalog, which is calibrated against Gaia DR1.

%\red{Add a table to describe the catalog columns. Add a figure to show some example light curves. }

%\red{Describe the organization of the released light curve data. }

\begin{figure*}[ht]
\centering
\includegraphics[width=\textwidth]{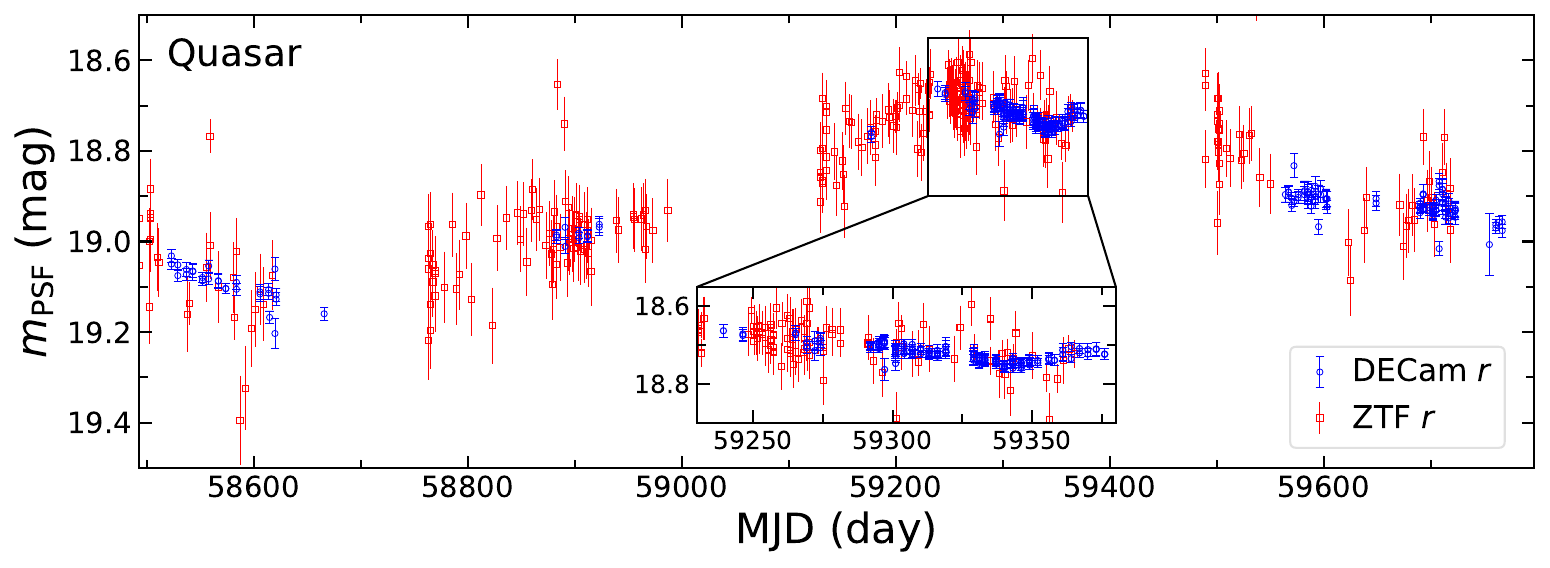}
\caption{Comparison of DECam $r$ band (open blue circle) and ZTF $r$ band (open red square) light curves of a quasar in the COSMOS field. The inset shows a zoom-in view around MJD$=$59300 day. The error bars indicate 1$\sigma$ uncertainty.}
\label{fig8}
\end{figure*}

\subsection{Example light curves}
Figure~\ref{fig7} shows example light curves of three RR Lyrae stars and three quasars in the COSMOS field. For RR Lyrae stars, we have modeled their light curves using \texttt{astropy} and presented phase-folded light curves. The high-quality photometry and dense time sampling ensure excellent recovery of period information. The overall clean phase-folded light curves of RR Lyrae stars also demonstrate the good performance of our zp calibration (Section~\ref{chk_cal}). The long time baseline, decent $\sim3$-day cadence, and multi-band coverage make it possible to qualitatively investigate both long and short timescale, as well as multi-band AGN variability. In particular, as shown in Figure~\ref{fig8} and its inset, the epoch magnitude uncertainty is $\sim5$ times smaller in HELM than in the ZTF DR19 -- median uncertainty of 0.014 mag from HELM $r$ band versus 0.068 mag from ZTF $r$ band. Therefore the resulting optical light curves from HELM are of much better quality than ZTF light curves, albeit with sparser cadences. 

\subsection{Caveats}

In this paper, our source catalogs for the CDF-S, ELAIS-S1, Stripe 82, and XMM-LSS fields are constructed based on the DES-DR2 source catalog, which has a median coadded catalog depth for a 1\farcs95 diameter aperture at SNR=10 of $g$=24.7, $r$=24.4, $i$=23.8, and $z$=23.1 mag. However, we may miss transients that were not detected in previous surveys. Given the number of exposures (up to 387 s) and total exposure time (up to 4480 s) in these fields, together with ongoing exposures that are not included in this data release, we would eventually achieve much deeper source detection by coadding all individual exposures. For the S-CVZ field, our source catalog is constructed based on catalogs of individual exposures. Therefore, the depth is limited by the deepest exposure. For the COSMOS field, we notice a source splitting issue for saturated objects in the HSC catalog.  As we are using deeper catalogs for source cross-matching (in particular for the COSMOS field), our source catalog may contain some spurious sources with low $N_{\rm exp}$ and close to or even below the typical depth of DECam individual exposures (relatively steep increase of source number toward the faint end in the $g$ and $i$ bands as seen in Figure~\ref{fig6}). Moreover, as the saturation magnitudes are quite faint for the HSC survey (18.0, 18.2, and 18.6 mag for the $gri$ bands; \citealp{HSC-SSP}), this issue may affect the sources that are above their saturation limits. Another major caveat is that there is no forced photometry or upper limits, which may give a false sense of the overall variability especially for sources close to the detection limit. The caveats mentioned above would be largely mitigated by building source catalogs directly from coadded DECam images and performing forced photometry on each source in our future data releases. On the other hand, any variable / transient sources that are not associated with a static object in the matched catalogs will be missed (e.g., off-axis transients and moving objects). These objects will be included in our future data releases after implementing dedicated pipeline.

\section{Summary}\label{sec:con}

In this work we present an overview of the HELM long-term monitoring program in extragalactic legacy fields, with the DECam imager on the 4m CTIO Blanco telescope. With a multi-year baseline and relatively high cadence, HELM aims to provide high-quality optical light curves for a large sample of AGNs in these fields, as part of the effort to obtain RM measurements for distant AGNs and quasars in the SDSS-V BHM-RM program. This long-term photometric monitoring program will also bridge past light curves and future LSST/WFST observations to compile decades-long light curves to facilitate investigations of long-term variability of astronomical objects (particularly AGNs). Given the ample multi-wavelength data in these legacy fields, the HELM light curve data will also provide legacy value to time-domain studies. At the time of writing, HELM has been extended to continue DECam monitoring in these fields through at least 2025B, given the delayed start of the LSST. 

In this work we also present the first data release from HELM, which includes source catalogs and light curves from the first 3.5 years of observations. In future data releases, we will provide fully calibrated individual and coadded images, as well as extended light curves.

\begin{deluxetable*}{llll}[ht]
\tablenum{3}
\caption{FITS table format for the median source catalog \label{table3}}
\tablehead{
\colhead{Column Name} & \colhead{Format} & \colhead{Units} & \colhead{Description}
}
\startdata
Index & LONG &  & Source index\\
ra & DOUBLE & degree & Right Ascension (J2000)\\
dec & DOUBLE & degree & Declination (J2000)\\
mag\_psf & DOUBLE & mag & \texttt{SExtractor} PSF magnitude\\
mag\_auto & DOUBLE & mag & \texttt{SExtractor} AUTO magnitude\\
flags & LONG &  & \texttt{SExtractor} extraction flags\\
spread\_model & DOUBLE & & \texttt{SExtractor} model-based star-galaxy classifier\\
class\_star & DOUBLE & & \texttt{SExtractor} neural network-based star-galaxy classifier\\
Nexp\_psf & LONG & & Number of exposures with successful mag\_psf measurement\\
Nexp\_auto & LONG & & Number of exposures with successful mag\_auto measurement\\
Separation & DOUBLE & arcsec & Distance between median coordinate in $griz$ bands and $u$ band (only in S-CVZ field)\\
\enddata
\tablecomments{All columns except Separation can have prefix, while all columns except Index, Nexp\_psf, Nexp\_auto, and Separation have suffix.\\
Prefix: ``*\_'' indicates the associated band for the parameter, with * = one or a combination of $ugriz$ bands (such as g\_ in all fields or griz\_ only for S-CVZ). If not specified (for Index, ra, and dec in all fields except S-CVZ), it represents the value for all bands combined.\\
Suffix: ``\_m'' and ``\_err'' indicate unweighted median (50\%) and its corresponding error of Nexp measurements with \_err $= 0.5 \times ({\rm 84th\ percentile - 16th\ percentile})/\sqrt{\rm Nexp}$; ``\_w'' indicates inverse variance weighted mean ($\Sigma(m_i/e_i^2)/\Sigma(1/e_i^2)$, where $m_i$ and $e_i$ represent individual measurement and its error at $i$th exposure, respectively); ''\_w\_err'' indicates the error associated with \_w after error propagation ($1/\sqrt{\Sigma(1/e_i^2)}$). Specifically, ``\_err'' = ``\_w\_err'' = error of individual measurement when Nexp=1.\\
Value = ``-99'' indicates no data or non-detection. }
\end{deluxetable*}

\begin{deluxetable*}{llll}[ht]
\tablenum{4}
\caption{FITS table format for the light curve \label{table4}}
\tablehead{
\colhead{Column Name} & \colhead{Format} & \colhead{Units} & \colhead{Description}
}
\startdata
ra & DOUBLE & degree & Right Ascension (J2000) \\
dec & DOUBLE & degree & Declination (J2000)\\
expnum & LONG & & DECam exposure number\\
mjd & DOUBLE & day & Modified Julian date of exposure\\
band & Char & & Filter (one of $ugriz$) \\
mag\_psf & DOUBLE & mag & \texttt{SExtractor} PSF magnitude\\
mag\_e\_psf & DOUBLE & mag & Uncertainty of \texttt{SExtractor} PSF magnitude\\
mag\_auto & DOUBLE & mag & \texttt{SExtractor} AUTO magnitude\\
mag\_e\_auto & DOUBLE & mag & Uncertainty of \texttt{SExtractor} AUTO magnitude\\
flags & LONG &  & \texttt{SExtractor} extraction flags\\
spread\_model & DOUBLE & & \texttt{SExtractor} model-based star-galaxy separation\\
spread\_model\_e & DOUBLE & & Uncertainty of \texttt{SExtractor} model-based star-galaxy separation\\
class\_star & DOUBLE & & \texttt{SExtractor} neural network-based star-galaxy separation\\
\enddata
\end{deluxetable*}

\begin{acknowledgments}
We thank Tamara M. Davis for initial discussions. This work is supported by NSF grants AST-2206499 and AST-2308077. F.E.B. acknowledges support from ANID - Millennium Science Initiative Program - ICN12\_009, CATA-BASAL - FB210003, and FONDECYT Regular - 1200495. Y.Q.X. acknowledges support from the National Key R\&D Program of China (2023YFA1608100), NSFC grants (12025303, 12393814), and the Strategic Priority Research Program of the Chinese Academy of Sciences (XDB0550300).
\end{acknowledgments}

\vspace{5mm}
\facilities{CTIO Blanco telescope}

%% Similar to \facility{}, there is the optional \software command to allow 
%% authors a place to specify which programs were used during the creation of 
%% the manuscript. Authors should list each code and include either a
%% citation or url to the code inside ()s when available.

\software{
\texttt{astropy} \citep{2013A&A...558A..33A, 2018AJ....156..123A},
\texttt{Matplotlib} \citep{Hunter2007}, 
\texttt{Numpy} \citep{Harris2020}, 
\texttt{SExtractor} \citep{1996A&AS..117..393B}}

%% Appendix material should be preceded with a single \appendix command.
%% There should be a \section command for each appendix. Mark appendix
%% subsections with the same markup you use in the main body of the paper.

%% Each Appendix (indicated with \section) will be lettered A, B, C, etc.
%% The equation counter will reset when it encounters the \appendix
%% command and will number appendix equations (A1), (A2), etc. The
%% Figure and Table counter will not reset.

%\appendix

\bibliography{sample631,refs}{}

\begin{thebibliography}{}
\expandafter\ifx\csname natexlab\endcsname\relax\def\natexlab#1{#1}\fi
\providecommand{\url}[1]{\href{#1}{#1}}
\providecommand{\dodoi}[1]{doi:~\href{http://doi.org/#1}{\nolinkurl{#1}}}
\providecommand{\doeprint}[1]{\href{http://ascl.net/#1}{\nolinkurl{http://ascl.net/#1}}}
\providecommand{\doarXiv}[1]{\href{https://arxiv.org/abs/#1}{\nolinkurl{https://arxiv.org/abs/#1}}}

\bibitem[{{Abbott} {et~al.}(2021){Abbott}, {Adam{\'o}w}, {Aguena}, {Allam},
  {Amon}, {Annis}, {Avila}, {Bacon}, {Banerji}, {Bechtol}, {Becker},
  {Bernstein}, {Bertin}, {Bhargava}, {Bridle}, {Brooks}, {Burke}, {Carnero
  Rosell}, {Carrasco Kind}, {Carretero}, {Castander}, {Cawthon}, {Chang},
  {Choi}, {Conselice}, {Costanzi}, {Crocce}, {da Costa}, {Davis}, {De Vicente},
  {DeRose}, {Desai}, {Diehl}, {Dietrich}, {Drlica-Wagner}, {Eckert},
  {Elvin-Poole}, {Everett}, {Evrard}, {Ferrero}, {Fert{\'e}}, {Flaugher},
  {Fosalba}, {Friedel}, {Frieman}, {Garc{\'\i}a-Bellido}, {Gaztanaga},
  {Gelman}, {Gerdes}, {Giannantonio}, {Gill}, {Gruen}, {Gruendl}, {Gschwend},
  {Gutierrez}, {Hartley}, {Hinton}, {Hollowood}, {Honscheid}, {Huterer},
  {James}, {Jeltema}, {Johnson}, {Kent}, {Kron}, {Kuehn}, {Kuropatkin},
  {Lahav}, {Li}, {Lidman}, {Lin}, {MacCrann}, {Maia}, {Manning}, {Maloney},
  {March}, {Marshall}, {Martini}, {Melchior}, {Menanteau}, {Miquel}, {Morgan},
  {Myles}, {Neilsen}, {Ogando}, {Palmese}, {Paz-Chinch{\'o}n}, {Petravick},
  {Pieres}, {Plazas}, {Pond}, {Rodriguez-Monroy}, {Romer}, {Roodman}, {Rykoff},
  {Sako}, {Sanchez}, {Santiago}, {Scarpine}, {Serrano}, {Sevilla-Noarbe},
  {Smith}, {Smith}, {Soares-Santos}, {Suchyta}, {Swanson}, {Tarle}, {Thomas},
  {To}, {Tremblay}, {Troxel}, {Tucker}, {Turner}, {Varga}, {Walker},
  {Wechsler}, {Weller}, {Wester}, {Wilkinson}, {Yanny}, {Zhang}, {Nikutta},
  {Fitzpatrick}, {Jacques}, {Scott}, {Olsen}, {Huang}, {Herrera}, {Juneau},
  {Nidever}, {Weaver}, {Adean}, {Correia}, {de Freitas}, {Freitas},
  {Singulani}, {Vila-Verde}, \& {Linea Science Server}}]{Abbott+2021ApJS}
{Abbott}, T.~M.~C., {Adam{\'o}w}, M., {Aguena}, M., {et~al.} 2021, \apjs, 255,
  20, \dodoi{10.3847/1538-4365/ac00b3}

\bibitem[{{Aihara} {et~al.}(2018){Aihara}, {Arimoto}, {Armstrong}, {Arnouts},
  {Bahcall}, {Bickerton}, {Bosch}, {Bundy}, {Capak}, {Chan}, {Chiba}, {Coupon},
  {Egami}, {Enoki}, {Finet}, {Fujimori}, {Fujimoto}, {Furusawa}, {Furusawa},
  {Goto}, {Goulding}, {Greco}, {Greene}, {Gunn}, {Hamana}, {Harikane},
  {Hashimoto}, {Hattori}, {Hayashi}, {Hayashi}, {He{\l}miniak}, {Higuchi},
  {Hikage}, {Ho}, {Hsieh}, {Huang}, {Huang}, {Ikeda}, {Imanishi}, {Inoue},
  {Iwasawa}, {Iwata}, {Jaelani}, {Jian}, {Kamata}, {Karoji}, {Kashikawa},
  {Katayama}, {Kawanomoto}, {Kayo}, {Koda}, {Koike}, {Kojima}, {Komiyama},
  {Konno}, {Koshida}, {Koyama}, {Kusakabe}, {Leauthaud}, {Lee}, {Lin}, {Lin},
  {Lupton}, {Mandelbaum}, {Matsuoka}, {Medezinski}, {Mineo}, {Miyama},
  {Miyatake}, {Miyazaki}, {Momose}, {More}, {More}, {Moritani}, {Moriya},
  {Morokuma}, {Mukae}, {Murata}, {Murayama}, {Nagao}, {Nakata}, {Niida},
  {Niikura}, {Nishizawa}, {Obuchi}, {Oguri}, {Oishi}, {Okabe}, {Okamoto},
  {Okura}, {Ono}, {Onodera}, {Onoue}, {Osato}, {Ouchi}, {Price}, {Pyo}, {Sako},
  {Sawicki}, {Shibuya}, {Shimasaku}, {Shimono}, {Shirasaki}, {Silverman},
  {Simet}, {Speagle}, {Spergel}, {Strauss}, {Sugahara}, {Sugiyama}, {Suto},
  {Suyu}, {Suzuki}, {Tait}, {Takada}, {Takata}, {Tamura}, {Tanaka}, {Tanaka},
  {Tanaka}, {Tanaka}, {Terai}, {Terashima}, {Toba}, {Tominaga}, {Toshikawa},
  {Turner}, {Uchida}, {Uchiyama}, {Umetsu}, {Uraguchi}, {Urata}, {Usuda},
  {Utsumi}, {Wang}, {Wang}, {Wong}, {Yabe}, {Yamada}, {Yamanoi}, {Yasuda},
  {Yeh}, {Yonehara}, \& {Yuma}}]{HSC-SSP}
{Aihara}, H., {Arimoto}, N., {Armstrong}, R., {et~al.} 2018, \pasj, 70, S4,
  \dodoi{10.1093/pasj/psx066}

\bibitem[{{Aihara} {et~al.}(2022){Aihara}, {AlSayyad}, {Ando}, {Armstrong},
  {Bosch}, {Egami}, {Furusawa}, {Furusawa}, {Harasawa}, {Harikane}, {Hsieh},
  {Ikeda}, {Ito}, {Iwata}, {Kodama}, {Koike}, {Kokubo}, {Komiyama}, {Li},
  {Liang}, {Lin}, {Lupton}, {Lust}, {MacArthur}, {Mawatari}, {Mineo},
  {Miyatake}, {Miyazaki}, {More}, {Morishima}, {Murayama}, {Nakajima},
  {Nakata}, {Nishizawa}, {Oguri}, {Okabe}, {Okura}, {Ono}, {Osato}, {Ouchi},
  {Pan}, {Plazas Malag{\'o}n}, {Price}, {Reed}, {Rykoff}, {Shibuya},
  {Simunovic}, {Strauss}, {Sugimori}, {Suto}, {Suzuki}, {Takada}, {Takagi},
  {Takata}, {Takita}, {Tanaka}, {Tang}, {Taranu}, {Terai}, {Toba}, {Turner},
  {Uchiyama}, {Vijarnwannaluk}, {Waters}, {Yamada}, {Yamamoto}, \&
  {Yamashita}}]{SSP-DR3}
{Aihara}, H., {AlSayyad}, Y., {Ando}, M., {et~al.} 2022, \pasj, 74, 247,
  \dodoi{10.1093/pasj/psab122}

\bibitem[{{Astropy Collaboration} {et~al.}(2013){Astropy Collaboration},
  {Robitaille}, {Tollerud}, {Greenfield}, {Droettboom}, {Bray}, {Aldcroft},
  {Davis}, {Ginsburg}, {Price-Whelan}, {Kerzendorf}, {Conley}, {Crighton},
  {Barbary}, {Muna}, {Ferguson}, {Grollier}, {Parikh}, {Nair}, {Unther},
  {Deil}, {Woillez}, {Conseil}, {Kramer}, {Turner}, {Singer}, {Fox}, {Weaver},
  {Zabalza}, {Edwards}, {Azalee Bostroem}, {Burke}, {Casey}, {Crawford},
  {Dencheva}, {Ely}, {Jenness}, {Labrie}, {Lim}, {Pierfederici}, {Pontzen},
  {Ptak}, {Refsdal}, {Servillat}, \& {Streicher}}]{2013A&A...558A..33A}
{Astropy Collaboration}, {Robitaille}, T.~P., {Tollerud}, E.~J., {et~al.} 2013,
  \aap, 558, A33, \dodoi{10.1051/0004-6361/201322068}

\bibitem[{{Astropy Collaboration} {et~al.}(2018){Astropy Collaboration},
  {Price-Whelan}, {Sip{\H{o}}cz}, {G{\"u}nther}, {Lim}, {Crawford}, {Conseil},
  {Shupe}, {Craig}, {Dencheva}, {Ginsburg}, {VanderPlas}, {Bradley},
  {P{\'e}rez-Su{\'a}rez}, {de Val-Borro}, {Aldcroft}, {Cruz}, {Robitaille},
  {Tollerud}, {Ardelean}, {Babej}, {Bach}, {Bachetti}, {Bakanov}, {Bamford},
  {Barentsen}, {Barmby}, {Baumbach}, {Berry}, {Biscani}, {Boquien}, {Bostroem},
  {Bouma}, {Brammer}, {Bray}, {Breytenbach}, {Buddelmeijer}, {Burke},
  {Calderone}, {Cano Rodr{\'\i}guez}, {Cara}, {Cardoso}, {Cheedella}, {Copin},
  {Corrales}, {Crichton}, {D'Avella}, {Deil}, {Depagne}, {Dietrich}, {Donath},
  {Droettboom}, {Earl}, {Erben}, {Fabbro}, {Ferreira}, {Finethy}, {Fox},
  {Garrison}, {Gibbons}, {Goldstein}, {Gommers}, {Greco}, {Greenfield},
  {Groener}, {Grollier}, {Hagen}, {Hirst}, {Homeier}, {Horton}, {Hosseinzadeh},
  {Hu}, {Hunkeler}, {Ivezi{\'c}}, {Jain}, {Jenness}, {Kanarek}, {Kendrew},
  {Kern}, {Kerzendorf}, {Khvalko}, {King}, {Kirkby}, {Kulkarni}, {Kumar},
  {Lee}, {Lenz}, {Littlefair}, {Ma}, {Macleod}, {Mastropietro}, {McCully},
  {Montagnac}, {Morris}, {Mueller}, {Mumford}, {Muna}, {Murphy}, {Nelson},
  {Nguyen}, {Ninan}, {N{\"o}the}, {Ogaz}, {Oh}, {Parejko}, {Parley}, {Pascual},
  {Patil}, {Patil}, {Plunkett}, {Prochaska}, {Rastogi}, {Reddy Janga},
  {Sabater}, {Sakurikar}, {Seifert}, {Sherbert}, {Sherwood-Taylor}, {Shih},
  {Sick}, {Silbiger}, {Singanamalla}, {Singer}, {Sladen}, {Sooley},
  {Sornarajah}, {Streicher}, {Teuben}, {Thomas}, {Tremblay}, {Turner},
  {Terr{\'o}n}, {van Kerkwijk}, {de la Vega}, {Watkins}, {Weaver}, {Whitmore},
  {Woillez}, {Zabalza}, \& {Astropy Contributors}}]{2018AJ....156..123A}
{Astropy Collaboration}, {Price-Whelan}, A.~M., {Sip{\H{o}}cz}, B.~M., {et~al.}
  2018, \aj, 156, 123, \dodoi{10.3847/1538-3881/aabc4f}

\bibitem[{{Bellm} {et~al.}(2019){Bellm}, {Kulkarni}, {Graham}, {Dekany},
  {Smith}, {Riddle}, {Masci}, {Helou}, {Prince}, {Adams}, {Barbarino},
  {Barlow}, {Bauer}, {Beck}, {Belicki}, {Biswas}, {Blagorodnova}, {Bodewits},
  {Bolin}, {Brinnel}, {Brooke}, {Bue}, {Bulla}, {Burruss}, {Cenko}, {Chang},
  {Connolly}, {Coughlin}, {Cromer}, {Cunningham}, {De}, {Delacroix}, {Desai},
  {Duev}, {Eadie}, {Farnham}, {Feeney}, {Feindt}, {Flynn}, {Franckowiak},
  {Frederick}, {Fremling}, {Gal-Yam}, {Gezari}, {Giomi}, {Goldstein},
  {Golkhou}, {Goobar}, {Groom}, {Hacopians}, {Hale}, {Henning}, {Ho}, {Hover},
  {Howell}, {Hung}, {Huppenkothen}, {Imel}, {Ip}, {Ivezi{\'c}}, {Jackson},
  {Jones}, {Juric}, {Kasliwal}, {Kaspi}, {Kaye}, {Kelley}, {Kowalski},
  {Kramer}, {Kupfer}, {Landry}, {Laher}, {Lee}, {Lin}, {Lin}, {Lunnan},
  {Giomi}, {Mahabal}, {Mao}, {Miller}, {Monkewitz}, {Murphy}, {Ngeow},
  {Nordin}, {Nugent}, {Ofek}, {Patterson}, {Penprase}, {Porter}, {Rauch},
  {Rebbapragada}, {Reiley}, {Rigault}, {Rodriguez}, {van Roestel}, {Rusholme},
  {van Santen}, {Schulze}, {Shupe}, {Singer}, {Soumagnac}, {Stein}, {Surace},
  {Sollerman}, {Szkody}, {Taddia}, {Terek}, {Van Sistine}, {van Velzen},
  {Vestrand}, {Walters}, {Ward}, {Ye}, {Yu}, {Yan}, \& {Zolkower}}]{ZTF}
{Bellm}, E.~C., {Kulkarni}, S.~R., {Graham}, M.~J., {et~al.} 2019, \pasp, 131,
  018002, \dodoi{10.1088/1538-3873/aaecbe}

\bibitem[{{Bernstein} {et~al.}(2012){Bernstein}, {Kessler}, {Kuhlmann},
  {Biswas}, {Kovacs}, {Aldering}, {Crane}, {D'Andrea}, {Finley}, {Frieman},
  {Hufford}, {Jarvis}, {Kim}, {Marriner}, {Mukherjee}, {Nichol}, {Nugent},
  {Parkinson}, {Reis}, {Sako}, {Spinka}, \& {Sullivan}}]{DES_supernova1}
{Bernstein}, J.~P., {Kessler}, R., {Kuhlmann}, S., {et~al.} 2012, \apj, 753,
  152, \dodoi{10.1088/0004-637X/753/2/152}

\bibitem[{{Bertin} \& {Arnouts}(1996)}]{1996A&AS..117..393B}
{Bertin}, E., \& {Arnouts}, S. 1996, \aaps, 117, 393,
  \dodoi{10.1051/aas:1996164}

\bibitem[{{Blandford} \& {McKee}(1982)}]{Blandford_McKee_1982}
{Blandford}, R.~D., \& {McKee}, C.~F. 1982, \apj, 255, 419,
  \dodoi{10.1086/159843}

\bibitem[{{Chambers} {et~al.}(2016){Chambers}, {Magnier}, {Metcalfe},
  {Flewelling}, {Huber}, {Waters}, {Denneau}, {Draper}, {Farrow}, {Finkbeiner},
  {Holmberg}, {Koppenhoefer}, {Price}, {Rest}, {Saglia}, {Schlafly}, {Smartt},
  {Sweeney}, {Wainscoat}, {Burgett}, {Chastel}, {Grav}, {Heasley}, {Hodapp},
  {Jedicke}, {Kaiser}, {Kudritzki}, {Luppino}, {Lupton}, {Monet}, {Morgan},
  {Onaka}, {Shiao}, {Stubbs}, {Tonry}, {White}, {Ba{\~n}ados}, {Bell},
  {Bender}, {Bernard}, {Boegner}, {Boffi}, {Botticella}, {Calamida},
  {Casertano}, {Chen}, {Chen}, {Cole}, {Deacon}, {Frenk}, {Fitzsimmons},
  {Gezari}, {Gibbs}, {Goessl}, {Goggia}, {Gourgue}, {Goldman}, {Grant},
  {Grebel}, {Hambly}, {Hasinger}, {Heavens}, {Heckman}, {Henderson}, {Henning},
  {Holman}, {Hopp}, {Ip}, {Isani}, {Jackson}, {Keyes}, {Koekemoer}, {Kotak},
  {Le}, {Liska}, {Long}, {Lucey}, {Liu}, {Martin}, {Masci}, {McLean}, {Mindel},
  {Misra}, {Morganson}, {Murphy}, {Obaika}, {Narayan}, {Nieto-Santisteban},
  {Norberg}, {Peacock}, {Pier}, {Postman}, {Primak}, {Rae}, {Rai}, {Riess},
  {Riffeser}, {Rix}, {R{\"o}ser}, {Russel}, {Rutz}, {Schilbach}, {Schultz},
  {Scolnic}, {Strolger}, {Szalay}, {Seitz}, {Small}, {Smith}, {Soderblom},
  {Taylor}, {Thomson}, {Taylor}, {Thakar}, {Thiel}, {Thilker}, {Unger},
  {Urata}, {Valenti}, {Wagner}, {Walder}, {Walter}, {Watters}, {Werner},
  {Wood-Vasey}, \& {Wyse}}]{PS1}
{Chambers}, K.~C., {Magnier}, E.~A., {Metcalfe}, N., {et~al.} 2016, arXiv
  e-prints, arXiv:1612.05560, \dodoi{10.48550/arXiv.1612.05560}

\bibitem[{{Drlica-Wagner} {et~al.}(2022){Drlica-Wagner}, {Ferguson},
  {Adam{\'o}w}, {Aguena}, {Allam}, {Andrade-Oliveira}, {Bacon}, {Bechtol},
  {Bell}, {Bertin}, {Bilaji}, {Bocquet}, {Bom}, {Brooks}, {Burke},
  {Carballo-Bello}, {Carlin}, {Carnero Rosell}, {Carrasco Kind}, {Carretero},
  {Castander}, {Cerny}, {Chang}, {Choi}, {Conselice}, {Costanzi},
  {Crnojevi{\'c}}, {da Costa}, {de Vicente}, {Desai}, {Esteves}, {Everett},
  {Ferrero}, {Fitzpatrick}, {Flaugher}, {Friedel}, {Frieman},
  {Garc{\'\i}a-Bellido}, {Gatti}, {Gaztanaga}, {Gerdes}, {Gruen}, {Gruendl},
  {Gschwend}, {Hartley}, {Hernandez-Lang}, {Hinton}, {Hollowood}, {Honscheid},
  {Hughes}, {Jacques}, {James}, {Johnson}, {Kuehn}, {Kuropatkin}, {Lahav},
  {Li}, {Lidman}, {Lin}, {March}, {Marshall}, {Mart{\'\i}nez-Delgado},
  {Mart{\'\i}nez-V{\'a}zquez}, {Massana}, {Mau}, {McNanna}, {Melchior},
  {Menanteau}, {Miller}, {Miquel}, {Mohr}, {Morgan}, {Mutlu-Pakdil},
  {Mu{\~n}oz}, {Neilsen}, {Nidever}, {Nikutta}, {Nilo Castellon}, {No{\"e}l},
  {Ogando}, {Olsen}, {Pace}, {Palmese}, {Paz-Chinch{\'o}n}, {Pereira},
  {Pieres}, {Plazas Malag{\'o}n}, {Prat}, {Riley}, {Rodriguez-Monroy}, {Romer},
  {Roodman}, {Sako}, {Sakowska}, {Sanchez}, {S{\'a}nchez}, {Sand},
  {Santana-Silva}, {Santiago}, {Schubnell}, {Serrano}, {Sevilla-Noarbe},
  {Simon}, {Smith}, {Soares-Santos}, {Stringfellow}, {Suchyta}, {Suson}, {Tan},
  {Tarle}, {Tavangar}, {Thomas}, {To}, {Tollerud}, {Troxel}, {Tucker}, {Varga},
  {Vivas}, {Walker}, {Weller}, {Wilkinson}, {Wu}, {Yanny}, {Zaborowski},
  {Zenteno}, {Delve Collaboration}, {Des Collaboration}, \& {Astro Data
  Lab}}]{Drlica-Wagner_etal_2022}
{Drlica-Wagner}, A., {Ferguson}, P.~S., {Adam{\'o}w}, M., {et~al.} 2022, \apjs,
  261, 38, \dodoi{10.3847/1538-4365/ac78eb}

\bibitem[{{Flaugher} {et~al.}(2015){Flaugher}, {Diehl}, {Honscheid}, {Abbott},
  {Alvarez}, {Angstadt}, {Annis}, {Antonik}, {Ballester}, {Beaufore},
  {Bernstein}, {Bernstein}, {Bigelow}, {Bonati}, {Boprie}, {Brooks},
  {Buckley-Geer}, {Campa}, {Cardiel-Sas}, {Castander}, {Castilla}, {Cease},
  {Cela-Ruiz}, {Chappa}, {Chi}, {Cooper}, {da Costa}, {Dede}, {Derylo},
  {DePoy}, {de Vicente}, {Doel}, {Drlica-Wagner}, {Eiting}, {Elliott}, {Emes},
  {Estrada}, {Fausti Neto}, {Finley}, {Flores}, {Frieman}, {Gerdes},
  {Gladders}, {Gregory}, {Gutierrez}, {Hao}, {Holland}, {Holm}, {Huffman},
  {Jackson}, {James}, {Jonas}, {Karcher}, {Karliner}, {Kent}, {Kessler},
  {Kozlovsky}, {Kron}, {Kubik}, {Kuehn}, {Kuhlmann}, {Kuk}, {Lahav}, {Lathrop},
  {Lee}, {Levi}, {Lewis}, {Li}, {Mandrichenko}, {Marshall}, {Martinez},
  {Merritt}, {Miquel}, {Mu{\~n}oz}, {Neilsen}, {Nichol}, {Nord}, {Ogando},
  {Olsen}, {Palaio}, {Patton}, {Peoples}, {Plazas}, {Rauch}, {Reil}, {Rheault},
  {Roe}, {Rogers}, {Roodman}, {Sanchez}, {Scarpine}, {Schindler}, {Schmidt},
  {Schmitt}, {Schubnell}, {Schultz}, {Schurter}, {Scott}, {Serrano}, {Shaw},
  {Smith}, {Soares-Santos}, {Stefanik}, {Stuermer}, {Suchyta}, {Sypniewski},
  {Tarle}, {Thaler}, {Tighe}, {Tran}, {Tucker}, {Walker}, {Wang}, {Watson},
  {Weaverdyck}, {Wester}, {Woods}, {Yanny}, \& {DES Collaboration}}]{DECam}
{Flaugher}, B., {Diehl}, H.~T., {Honscheid}, K., {et~al.} 2015, \aj, 150, 150,
  \dodoi{10.1088/0004-6256/150/5/150}

\bibitem[{{Graham} {et~al.}(2015){Graham}, {Djorgovski}, {Stern}, {Drake},
  {Mahabal}, {Donalek}, {Glikman}, {Larson}, \&
  {Christensen}}]{search_binary_BH}
{Graham}, M.~J., {Djorgovski}, S.~G., {Stern}, D., {et~al.} 2015, \mnras, 453,
  1562, \dodoi{10.1093/mnras/stv1726}

\bibitem[{{Graham} {et~al.}(2023){Graham}, {Knop}, {Kennedy}, {Nugent},
  {Bellm}, {Catelan}, {Patel}, {Smotherman}, {Soraisam}, {Stetzler},
  {Aldoroty}, {Awbrey}, {Baeza-Villagra}, {Bernardinelli}, {Bianco}, {Brout},
  {Clarke}, {Clarkson}, {Collett}, {Davenport}, {Fu}, {Gizis}, {Heinze}, {Hu},
  {Jha}, {Juri{\'c}}, {Kalmbach}, {Kim}, {Lee}, {Lidman}, {Magee},
  {Mart{\'\i}nez-V{\'a}zquez}, {Matheson}, {Narayan}, {Palmese}, {Phillips},
  {Rabus}, {Rest}, {Rodr{\'\i}guez-Segovia}, {Street}, {Vivas}, {Wang}, {Wolf},
  \& {Yang}}]{Graham_etal_2023}
{Graham}, M.~L., {Knop}, R.~A., {Kennedy}, T.~D., {et~al.} 2023, \mnras, 519,
  3881, \dodoi{10.1093/mnras/stac3363}

\bibitem[{Harris {et~al.}(2020)Harris, Millman, van~der Walt, Gommers,
  Virtanen, Cournapeau, Wieser, Taylor, Berg, Smith, Kern, Picus, Hoyer, van
  Kerkwijk, Brett, Haldane, del R{\'{i}}o, Wiebe, Peterson,
  G{\'{e}}rard-Marchant, Sheppard, Reddy, Weckesser, Abbasi, Gohlke, \&
  Oliphant}]{Harris2020}
Harris, C.~R., Millman, K.~J., van~der Walt, S.~J., {et~al.} 2020, Nature, 585,
  357, \dodoi{10.1038/s41586-020-2649-2}

\bibitem[{{Henden} {et~al.}(2015){Henden}, {Levine}, {Terrell}, \&
  {Welch}}]{APASSDR9}
{Henden}, A.~A., {Levine}, S., {Terrell}, D., \& {Welch}, D.~L. 2015, in
  American Astronomical Society Meeting Abstracts, Vol. 225, American
  Astronomical Society Meeting Abstracts \#225, 336.16

\bibitem[{Hunter(2007)}]{Hunter2007}
Hunter, J.~D. 2007, Computing in Science \& Engineering, 9, 90,
  \dodoi{10.1109/MCSE.2007.55}

\bibitem[{{Jones} {et~al.}(2017){Jones}, {Scolnic}, {Riess}, {Kessler}, {Rest},
  {Kirshner}, {Berger}, {Ortega}, {Foley}, {Chornock}, {Challis}, {Burgett},
  {Chambers}, {Draper}, {Flewelling}, {Huber}, {Kaiser}, {Kudritzki},
  {Metcalfe}, {Wainscoat}, \& {Waters}}]{PS1_MDS}
{Jones}, D.~O., {Scolnic}, D.~M., {Riess}, A.~G., {et~al.} 2017, \apj, 843, 6,
  \dodoi{10.3847/1538-4357/aa767b}

\bibitem[{{Kollmeier} {et~al.}(2017){Kollmeier}, {Zasowski}, {Rix}, {Johns},
  {Anderson}, {Drory}, {Johnson}, {Pogge}, {Bird}, {Blanc}, {Brownstein},
  {Crane}, {De Lee}, {Klaene}, {Kreckel}, {MacDonald}, {Merloni}, {Ness},
  {O'Brien}, {Sanchez-Gallego}, {Sayres}, {Shen}, {Thakar}, {Tkachenko},
  {Aerts}, {Blanton}, {Eisenstein}, {Holtzman}, {Maoz}, {Nandra}, {Rockosi},
  {Weinberg}, {Bovy}, {Casey}, {Chaname}, {Clerc}, {Conroy}, {Eracleous},
  {G{\"a}nsicke}, {Hekker}, {Horne}, {Kauffmann}, {McQuinn}, {Pellegrini},
  {Schinnerer}, {Schlafly}, {Schwope}, {Seibert}, {Teske}, \& {van
  Saders}}]{Kollmeier_etal_2017}
{Kollmeier}, J.~A., {Zasowski}, G., {Rix}, H.-W., {et~al.} 2017, ArXiv
  e-prints.
\newblock \doarXiv{1711.03234}

\bibitem[{{Lacy} {et~al.}(2021){Lacy}, {Surace}, {Farrah}, {Nyland}, {Afonso},
  {Brandt}, {Clements}, {Lagos}, {Maraston}, {Pforr}, {Sajina}, {Sako},
  {Vaccari}, {Wilson}, {Ballantyne}, {Barkhouse}, {Brunner}, {Cane}, {Clarke},
  {Cooper}, {Cooray}, {Covone}, {D'Andrea}, {Evrard}, {Ferguson}, {Frieman},
  {Gonzalez-Perez}, {Gupta}, {Hatziminaoglou}, {Huang}, {Jagannathan},
  {Jarvis}, {Jones}, {Kimball}, {Lidman}, {Lubin}, {Marchetti}, {Martini},
  {McMahon}, {Mei}, {Messias}, {Murphy}, {Newman}, {Nichol}, {Norris},
  {Oliver}, {Perez-Fournon}, {Peters}, {Pierre}, {Polisensky}, {Richards},
  {Ridgway}, {R{\"o}ttgering}, {Seymour}, {Shirley}, {Somerville}, {Strauss},
  {Suntzeff}, {Thorman}, {van Kampen}, {Verma}, {Wechsler}, \&
  {Wood-Vasey}}]{Lacy+2021MNRAS}
{Lacy}, M., {Surace}, J.~A., {Farrah}, D., {et~al.} 2021, \mnras, 501, 892,
  \dodoi{10.1093/mnras/staa3714}

\bibitem[{{Morganson} {et~al.}(2018){Morganson}, {Gruendl}, {Menanteau},
  {Carrasco Kind}, {Chen}, {Daues}, {Drlica-Wagner}, {Friedel}, {Gower},
  {Johnson}, {Johnson}, {Kessler}, {Paz-Chinch{\'o}n}, {Petravick}, {Pond},
  {Yanny}, {Allam}, {Armstrong}, {Barkhouse}, {Bechtol}, {Benoit-L{\'e}vy},
  {Bernstein}, {Bertin}, {Buckley-Geer}, {Covarrubias}, {Desai}, {Diehl},
  {Goldstein}, {Gruen}, {Li}, {Lin}, {Marriner}, {Mohr}, {Neilsen}, {Ngeow},
  {Paech}, {Rykoff}, {Sako}, {Sevilla-Noarbe}, {Sheldon}, {Sobreira}, {Tucker},
  {Wester}, \& {DES Collaboration}}]{Morganson_etal_2018}
{Morganson}, E., {Gruendl}, R.~A., {Menanteau}, F., {et~al.} 2018, \pasp, 130,
  074501, \dodoi{10.1088/1538-3873/aab4ef}

\bibitem[{{Peterson}(1993)}]{Peterson_1993}
{Peterson}, B.~M. 1993, \pasp, 105, 247, \dodoi{10.1086/133140}

\bibitem[{{Sako} {et~al.}(2018){Sako}, {Bassett}, {Becker}, {Brown},
  {Campbell}, {Wolf}, {Cinabro}, {D'Andrea}, {Dawson}, {DeJongh}, {Depoy},
  {Dilday}, {Doi}, {Filippenko}, {Fischer}, {Foley}, {Frieman}, {Galbany},
  {Garnavich}, {Goobar}, {Gupta}, {Hill}, {Hayden}, {Hlozek}, {Holtzman},
  {Hopp}, {Jha}, {Kessler}, {Kollatschny}, {Leloudas}, {Marriner}, {Marshall},
  {Miquel}, {Morokuma}, {Mosher}, {Nichol}, {Nordin}, {Olmstead}, {{\"O}stman},
  {Prieto}, {Richmond}, {Romani}, {Sollerman}, {Stritzinger}, {Schneider},
  {Smith}, {Wheeler}, {Yasuda}, \& {Zheng}}]{SDSS_supernova}
{Sako}, M., {Bassett}, B., {Becker}, A.~C., {et~al.} 2018, \pasp, 130, 064002,
  \dodoi{10.1088/1538-3873/aab4e0}

\bibitem[{{Shen} {et~al.}(2015){Shen}, {Brandt}, {Dawson}, {Hall}, {McGreer},
  {Anderson}, {Chen}, {Denney}, {Eftekharzadeh}, {Fan}, {Gao}, {Green},
  {Greene}, {Ho}, {Horne}, {Jiang}, {Kelly}, {Kinemuchi}, {Kochanek},
  {P{\^a}ris}, {Peters}, {Peterson}, {Petitjean}, {Ponder}, {Richards},
  {Schneider}, {Seth}, {Smith}, {Strauss}, {Tao}, {Trump}, {Wood-Vasey}, {Zu},
  {Eisenstein}, {Pan}, {Bizyaev}, {Malanushenko}, {Malanushenko}, \&
  {Oravetz}}]{Shen_etal_2015a}
{Shen}, Y., {Brandt}, W.~N., {Dawson}, K.~S., {et~al.} 2015, \apjs, 216, 4,
  \dodoi{10.1088/0067-0049/216/1/4}

\bibitem[{{Smith} {et~al.}(2020){Smith}, {D'Andrea}, {Sullivan}, {M{\"o}ller},
  {Nichol}, {Thomas}, {Kim}, {Sako}, {Castander}, {Filippenko}, {Foley},
  {Galbany}, {Gonz{\'a}lez-Gait{\'a}n}, {Kasai}, {Kirshner}, {Lidman},
  {Scolnic}, {Brout}, {Davis}, {Gupta}, {Hinton}, {Kessler}, {Lasker},
  {Macaulay}, {Wolf}, {Zhang}, {Asorey}, {Avelino}, {Bassett}, {Calcino},
  {Carollo}, {Casas}, {Challis}, {Childress}, {Clocchiatti}, {Crawford},
  {Frohmaier}, {Glazebrook}, {Goldstein}, {Graham}, {Hoormann}, {Kuehn},
  {Lewis}, {Mandel}, {Morganson}, {Muthukrishna}, {Nugent}, {Pan},
  {Pursiainen}, {Sharp}, {Sommer}, {Swann}, {Thomas}, {Tucker}, {Uddin},
  {Wiseman}, {Zheng}, {Abbott}, {Annis}, {Avila}, {Bechtol}, {Bernstein},
  {Bertin}, {Brooks}, {Burke}, {Carnero Rosell}, {Carrasco Kind}, {Carretero},
  {Cunha}, {da Costa}, {Davis}, {De Vicente}, {Diehl}, {Eifler}, {Estrada},
  {Frieman}, {Garc{\'\i}a-Bellido}, {Gaztanaga}, {Gerdes}, {Gruen}, {Gruendl},
  {Gschwend}, {Gutierrez}, {Hartley}, {Hollowood}, {Honscheid}, {Hoyle},
  {James}, {Johnson}, {Johnson}, {Kuropatkin}, {Li}, {Lima}, {Maia}, {March},
  {Marshall}, {Martini}, {Menanteau}, {Miller}, {Miquel}, {Neilsen}, {Ogando},
  {Plazas}, {Romer}, {Sanchez}, {Scarpine}, {Schubnell}, {Serrano},
  {Sevilla-Noarbe}, {Soares-Santos}, {Sobreira}, {Suchyta}, {Tarle}, {Tucker},
  \& {Wester}}]{DES_supernova2}
{Smith}, M., {D'Andrea}, C.~B., {Sullivan}, M., {et~al.} 2020, \aj, 160, 267,
  \dodoi{10.3847/1538-3881/abc01b}

\bibitem[{{Wang} {et~al.}(2023){Wang}, {Liu}, {Cai}, {Geng}, {Fang}, {He},
  {Jiang}, {Jiang}, {Kong}, {Li}, {Li}, {Luo}, {Pan}, {Wu}, {Yang}, {Yu},
  {Zheng}, {Zhu}, {Cai}, {Chen}, {Chen}, {Dai}, {Fan}, {Fan}, {Fang}, {He},
  {Hu}, {Hu}, {Jin}, {Jiang}, {Li}, {Li}, {Li}, {Liang}, {Lin}, {Liu}, {Liu},
  {Liu}, {Liu}, {Liu}, {Lou}, {Qu}, {Sheng}, {Shi}, {Shu}, {Su}, {Sun}, {Wang},
  {Wang}, {Wang}, {Wang}, {Wei}, {Wei}, {Xue}, {Yan}, {Yang}, {Yuan}, {Yuan},
  {Zhang}, {Zhang}, {Zhao}, \& {Zhao}}]{WFST}
{Wang}, T., {Liu}, G., {Cai}, Z., {et~al.} 2023, Science China Physics,
  Mechanics, and Astronomy, 66, 109512, \dodoi{10.1007/s11433-023-2197-5}

\bibitem[{{York} {et~al.}(2000){York}, {Adelman}, {Anderson}, {Anderson},
  {Annis}, {Bahcall}, {Bakken}, {Barkhouser}, {Bastian}, {Berman}, {Boroski},
  {Bracker}, {Briegel}, {Briggs}, {Brinkmann}, {Brunner}, {Burles}, {Carey},
  {Carr}, {Castander}, {Chen}, {Colestock}, {Connolly}, {Crocker}, {Csabai},
  {Czarapata}, {Davis}, {Doi}, {Dombeck}, {Eisenstein}, {Ellman}, {Elms},
  {Evans}, {Fan}, {Federwitz}, {Fiscelli}, {Friedman}, {Frieman}, {Fukugita},
  {Gillespie}, {Gunn}, {Gurbani}, {de Haas}, {Haldeman}, {Harris}, {Hayes},
  {Heckman}, {Hennessy}, {Hindsley}, {Holm}, {Holmgren}, {Huang}, {Hull},
  {Husby}, {Ichikawa}, {Ichikawa}, {Ivezi{\'c}}, {Kent}, {Kim}, {Kinney},
  {Klaene}, {Kleinman}, {Kleinman}, {Knapp}, {Korienek}, {Kron}, {Kunszt},
  {Lamb}, {Lee}, {Leger}, {Limmongkol}, {Lindenmeyer}, {Long}, {Loomis},
  {Loveday}, {Lucinio}, {Lupton}, {MacKinnon}, {Mannery}, {Mantsch}, {Margon},
  {McGehee}, {McKay}, {Meiksin}, {Merelli}, {Monet}, {Munn}, {Narayanan},
  {Nash}, {Neilsen}, {Neswold}, {Newberg}, {Nichol}, {Nicinski}, {Nonino},
  {Okada}, {Okamura}, {Ostriker}, {Owen}, {Pauls}, {Peoples}, {Peterson},
  {Petravick}, {Pier}, {Pope}, {Pordes}, {Prosapio}, {Rechenmacher}, {Quinn},
  {Richards}, {Richmond}, {Rivetta}, {Rockosi}, {Ruthmansdorfer}, {Sandford},
  {Schlegel}, {Schneider}, {Sekiguchi}, {Sergey}, {Shimasaku}, {Siegmund},
  {Smee}, {Smith}, {Snedden}, {Stone}, {Stoughton}, {Strauss}, {Stubbs},
  {SubbaRao}, {Szalay}, {Szapudi}, {Szokoly}, {Thakar}, {Tremonti}, {Tucker},
  {Uomoto}, {Vanden Berk}, {Vogeley}, {Waddell}, {Wang}, {Watanabe},
  {Weinberg}, {Yanny}, {Yasuda}, \& {SDSS Collaboration}}]{SDSS}
{York}, D.~G., {Adelman}, J., {Anderson}, John~E., J., {et~al.} 2000, \aj, 120,
  1579, \dodoi{10.1086/301513}

\bibitem[{{Zou} {et~al.}(2022){Zou}, {Brandt}, {Chen}, {Leja}, {Ni}, {Yan},
  {Yang}, {Zhu}, {Luo}, {Nyland}, {Vito}, \& {Xue}}]{Zou+2022ApJS}
{Zou}, F., {Brandt}, W.~N., {Chen}, C.-T., {et~al.} 2022, \apjs, 262, 15,
  \dodoi{10.3847/1538-4365/ac7bdf}

\end{thebibliography}
\bibliographystyle{aasjournal}

%% This command is needed to show the entire author+affiliation list when
%% the collaboration and author truncation commands are used.  It has to
%% go at the end of the manuscript.
%\allauthors

%% Include this line if you are using the \added, \replaced, \deleted
%% commands to see a summary list of all changes at the end of the article.
%\listofchanges
\end{CJK*}
\end{document}